\documentclass{aa}  

\usepackage{natbib}
\bibpunct{(}{)}{;}{a}{}{,}  
\usepackage{amsmath}
\usepackage{graphicx}
\usepackage{longtable}
\usepackage[varg]{txfonts}
\begin{document}
   \title{Quest for finding the lost siblings of the Sun 
   		\thanks{Based on observations made with Nordic Optical Telescope at La Palma under programme 44-014. Based on observations made with ESO VLT Kueyen Telescope at the Paranal observatory under program me ID 085.C-0062(A), 087.D-0010(A), and 088.B-0820(A). Based on data obtained from the ESO Science Archive Facility under programme ID 078.D-0080(A), 080,A-9006(A), 082.C-0446(A), 082.A-9007(A), 083,A-9004(B), and 089.C-0524(A).} \fnmsep
		\thanks{Full Table~2 is only available in electronic form at the CDS.}}


   \author{C. Liu
   	\inst{1}
          \and
          G. Ruchti\inst{1}
          \and
          S. Feltzing\inst{1}
          \and
          C. A. Mart\'inez-Barbosa\inst{2}
          \and
          T. Bensby\inst{1}
          \and
          A. G. A. Brown\inst{2}
          \and
          S. F. Portegies Zwart\inst{2}
          }

   \institute{Lund Observatory, Department of Astronomy and Theoretical Physics,
              Box 43, SE--221 00 Lund, Sweden\\
              \email{[cheng, greg, sofia]@astro.lu.se}
              \and
              Leiden Observatory, University of Leiden, P.B. 9513, Leiden 2300 RA, the Netherlands\\
              }

   \date{Received: 16 September 2014; accepted: 19 November 2014}

 
  \abstract
   {}
   {The aim of this paper is to find lost siblings of the Sun by analyzing high resolution spectra. Finding solar siblings will enable us to constrain the parameters of the parental cluster and the birth place of the Sun in the Galaxy.}
   {The solar siblings can be identified by accurate measurements of metallicity, stellar age and elemental abundances for solar neighbourhood stars. The solar siblings candidates were kinematically selected based on their proper motions, parallaxes and colours. Stellar parameters were determined through a purely spectroscopic approach and partly physical method, respectively. Comparing synthetic with observed spectra, elemental abundances were computed based on the stellar parameters obtained using a partly physical method. A chemical tagging technique was used to identify the solar siblings.}
   {We present stellar parameters, stellar ages, and detailed elemental abundances for Na, Mg, Al, Si, Ca, Ti, Cr, Fe, and Ni for 32 solar sibling candidates. Our abundances analysis shows that four stars are chemically homogenous together with the Sun. Technique of chemical tagging gives us a high probability that they might be from the same open cluster. Only one candidate --HIP 40317-- which has solar metallicity and age could be a solar sibling. We performed simulations of the Sun's birth cluster in analytical Galactic model and found that most of the radial velocities of the solar siblings lie in the range $-10 \leq \mathrm{V_r}\leq 10$ $\mathrm{km~s^{-1}}$, which is smaller than the radial velocity of HIP 40317 $(\mathrm{V_r} = 34.2~\mathrm{km~s^{-1}})$, under different Galactic parameters and different initial conditions of the Sun's birth cluster. The sibling status for HIP 40317 is not directly supported by our dynamical analysis.}
   {}


   \maketitle
%

\section{Introduction}

It is commonly thought that stars are born within clusters and the majority of embedded clusters do not survive longer than 10 Myr \citep{2003ARA&A..41...57L}. The Sun, like most stars, could be born in a cluster. The stars that were born together with the Sun are called solar siblings. Finding siblings of the Sun will enable us to determine the birth place of the Sun \citep{2009ApJ...696L..13P}, and to better understand the mechanisms of radial migration in the Galactic disk (Minchev \& Famaey 2010). It should be possible to identify them by obtaining accurate measurements of their kinematics, metallicities, elemental abundances and ages. 

Based on implications on the formation and morphology of the solar system and presence of short lived radioactive nuclei in meteorites, the possible birth environment of the Sun has been probed by several studies. As discussed by \cite{2009ApJ...696L..13P}, parent cluster could contain $10^{3} - 10^{4}$ stars and the size of the proto-Solar-cluster was between 0.5 and 3 pc. The same characters of parent cluster was found by \cite{2010ARA&A..48...47A}. He also pointed out that a massive supernova explosion happened about 0.1 -- 0.3 pc from the Sun. \cite{1996A&A...314..438W} found that the Sun travel outward by about 2 kpc from the birth place over the past 4.6 billion years. Comparing the cosmic abundance standard which was obtained by measuring early B-type stars in solar neighbourhood with the solar standard, \cite{2012A&A...539A.143N} also claimed that the Sun has migrated outward from its birth place in the inner disk at 5--6 kpc Galactic distance over its lifetime to current position. The radial migration of stars might be caused by transient spiral arms at corotation: churning \citep{2002MNRAS.336..785S}.

Using simulations \cite{2009ApJ...696L..13P} found that about 10--60 solar siblings could still be within 100 pc of the Sun assuming the parental cluster consisting of $\sim10^{3}$ stars. Following the previous analysis, \cite{2010MNRAS.407..458B} simulated the orbits of the stars in the Sun's birth cluster rather than tracing back the Sun's orbit over the whole lifetime in an analytic Galactic potential. The first potential candidate as a solar sibling was found based on their simulated phase-space distribution of the siblings. By taking into account the perturbation from spiral arms in Galactic potential, \cite{2011AstL...37..550B} found two interesting stars by constructing their Galactic orbits and analysing the parameters of encounter with solar orbit. Another potential candidate was found by considering the chemical compositions, age, and kinematics properties of FGK stars from Geneva-Copenhagen Survey Catalogues \citep{2009A&A...501..941H} by \cite{2012NewA...17..514B}. One more potential candidate was found in a search for solar siblings using the HARPS \citep{2014A&A...564A..43B}. However, \cite{2011MNRAS.412.1771M} argued that the solar siblings are unlikely to be found within 100 pc from the Sun, because an unbound open cluster is dispersed in a short period of time under the perturbation of the spiral gravitational field. Then, the members of cluster are scattered over a very large portion of the Galactic disc after 4.6 Gyr of dynamical evolution. In addition to the radial migration, the original kinematical information has been modified as a result of perturbations of spiral arms and central bar. They also pointed out that we still have chance to find the solar siblings in the solar vicinity if the parental cluster has $\sim10^{4}$ stars \citep{2011MNRAS.412.1771M}. The first real solar sibling HD 162826 which satisfies both chemical and strictly dynamical conditions was found by \cite{2014ApJ...787..154R}. This is encouraging and strengthens our ability to find the lost siblings of the Sun.

Since the original kinematical information of a star may be lost under the Galactic dynamic evolution, it will not be the first option to identify the solar siblings. The chemical information on the other hand is preserved in the form of elemental abundances in individual stars. In order to reconstruct dissolved star clusters, \cite{2002ARA&A..40..487F} first proposed the technique of chemical tagging based on understanding of chemical signatures that members of an open cluster are chemically homogeneous. The homogeneity of abundances in open clusters and moving groups have been demonstrated by recent studies \citep{2007AJ....133..694D, 2007AJ....133.1161D, 2009PASA...26...11D, 2010A&A...511A..56P}. A pair-wise metric, quantifying differences in chemical signatures between different clusters and the stars within a given cluster, has been defined by \cite{2013MNRAS.428.2321M}. This metric was applied to more than 30 open clusters with good measurement of elemental abundances, and they found that it is effective ($\geqslant$ 9\% of the total sample of stars, see also \citealt{2014MNRAS.438.2753M}) in detecting the members of clusters.

In Section 2, we present our sample of sibling candidates. We describe observations and the process of data reductions in Section 3. Both stellar parameters and elemental abundances are determined in Section 4. Our algorithm of chemical tagging is explained in Section 5. In Section 6, we give constraints on metallicity, stellar age and elemental abundances and find only one candidates could be solar sibling in our sample, and in Section 7 we give dynamical analysis on former identified star. Finally, conclusions are drawn in Section 8.


\section{Selecting solar sibling candidates}
\label{sect:sel}

\begin{figure}[htbp]
\begin{center}
\includegraphics[scale=0.5]{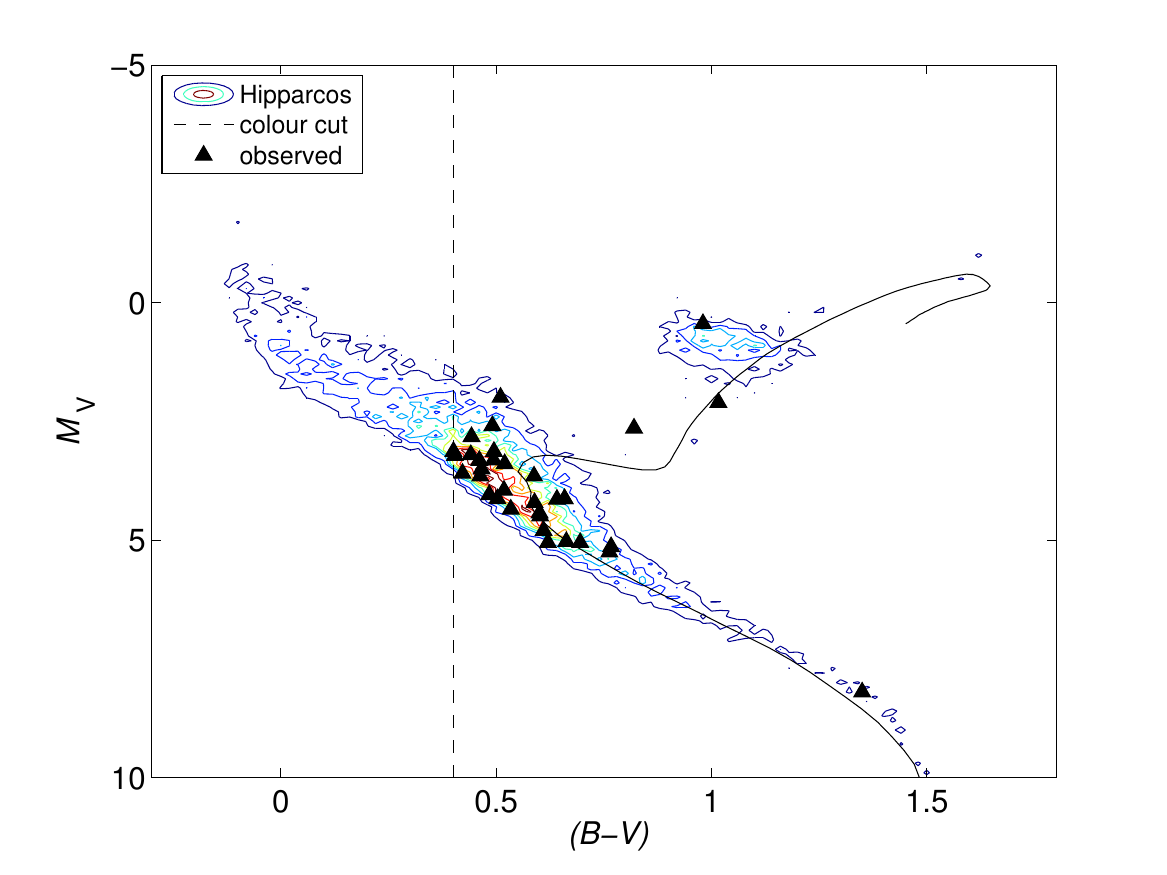}
\caption{Colour-magnitude diagram showing the absolute magnitude $M_{\rm V}$ versus. $B-V$. The contour plot shows the distribution of the stars in the Hipparcos catalogue with $\sigma_{\varpi} / \varpi \leq 0.1$ and $\sigma_{\rm B-V} \leq 0.05$ \citep{2007A&A...474..653V}. The spectra of the 33 sibling candidates are indicated with triangles. The solid line shows an isochrone with the age (5 Gyr) and metallicity of the Sun \citep{2004ApJS..155..667D}. The vertical dashed line indicate our cut-off at $B-V =$ 0.4.}
\label{cancmd}
\end{center}
\end{figure}

Solar sibling candidates were selected following the same methods and steps as in \citet{2010MNRAS.407..458B}. Assuming solar siblings have almost the same orbit as the Sun and taking the varying distance into account, the upper limit on the proper motion value can be obtained in their simulations. Stars within 100 pc from the Sun are selected. The predicted proper motion versus parallax phase-space was used as a first selection of solar sibling candidates in the Hipparcos Catalogue \citep{2007A&A...474..653V}. The exact selection criteria were: 
\begin{equation}
\varpi \geq 10 \ {\rm mas} \wedge \sigma_{\varpi} / \varpi \leq 0.1 \wedge \mu \leq 6.5 \ \rm mas \ yr^{-1},
\label{sc}
\end{equation}
where $\varpi$ and $\mu$ are the parallax and proper motion of the stars, respectively, and $\sigma_{\varpi}$ is the precision of the parallax. Since the Sun and solar siblings form together in the parent cluster, the siblings should have about the same age as the Sun. Inspection of stellar isochrones show that for solar metallicity a star with a colour of $B-V \le$ 0.4 is too young to be a solar sibling. Finally, 57 candidates were selected from the Hipparcos catalogue using two constraints. In this work, high resolution spectra of 33 of 57 candidates are analysed. Basic properties of the 33 sibling candidates, shown in Fig.~\ref{cancmd}, were collected from the Str$\ddot{\rm o}$mgren photometric and Hipparcos catalogues \citep{2007A&A...474..653V, 1983A&AS...54...55O, 1994A&AS..106..257O}, respectively. The data are listed in Table~\ref{canp}.

\begin{table*}
\begin{center}
\setlength{\tabcolsep}{8pt}
\begin{tiny}
\caption{Basic data for our solar sibling candidates.}
\label{canp}
\begin{tabular}{lllrrrrrrrrrrrrrrc}
\hline\hline
HIP&$V$&$\varpi$&$\mu_{\alpha \cos\delta}$&$\mu_{\delta}$&$B-V$&$b-y$&$c_{1}$&flag&$v{\rm \sin}i$&Instrument\\
       &(mag)&(mas)&(mas yr$^{-1}$)&(mas yr$^{-1}$)&(mag)&(mag)&(mag)&&($\mathrm{km~s}^{-1}$)&\\
\hline
7764      &8.83   &15.60 &0.13 &4.69  &0.610  &  &  &  &2.2  &FIES\\
8444     &8.76   &13.14 &1.06 &--1.98 &0.534  &  &  &  &7.4  &UVES\\
9405     &8.25   &11.22 &0.33 &4.63 &0.466  &  &   &   &5.9  &FIES, UVES\\
10786   &8.71   &14.33 &0.82 &1.56 &0.602  &  &  &  &20.4  &FEROS\\
14640   &7.62   &13.08 &--4.95 &0.57 &0.441  &0.306  &0.477  &$b$   &15.4  &FIES, UVES\\
15929   &8.44   &13.77 &0.55 &--2.24 &0.503  &0.313  &0.367  &$b$   &51.0  &FIES\\
21158   &7.08   &25.84 &--1.88 &--4.05 &0.641  &0.394  &0.370  &$b$  &1.9  &FIES\\
22002   &9.31  &14.69 &1.21 &--3.48 &0.602  &  &   &  &18.3  &FEROS\\
24232   &7.64   &13.71 &2.53 &1.82 &0.461  &0.307  &0.463  &$a$   &14.5  &FIES\\
25358   &8.19  &10.98 &0.70 &0.82 &0.520  &  &  &  &26.3  &FEROS\\
26744   &6.74   &14.81 &0.49 &--0.18 &0.491  &0.314  &0.492  &$a$   &8.9  &FIES\\
30344   &7.37   &34.10 &0.62 &--2.87 &0.663  &0.419  &0.317  &$b$  &16.0    &FEROS\\
33275   &7.60   &18.61 &0.75 &--0.81 &0.518  &0.334  &0.384  &$b$   &6.3  &FIES\\
33685   &7.39   &14.67 &4.93 &--1.18 &0.404  &0.268  &0.468  &$b$  &8.9  &FIES\\
40317   &9.58    &12.43 &--2.26 &2.87 &0.695  &  &  &  &1.7  &FIES, UVES\\
48062   &8.58   &12.34 &0.45 &1.90 &0.484  &  &  &  &22.7  &FIES, UVES\\
51581   &8.43   &11.05 &0.84 &2.94 &0.588  &0.366  &0.391  &$a$  &6.3  &UVES\\
53921   &7.25   &15.01 &0.48 &--3.37 &0.495  &0.314  &0.439  &$b$  &5.3  &UVES\\
56798   &8.73    &12.05 &--4.72 &--3.13 &0.659  &0.413  &0.385  &$b$  &1.9  &FIES\\
58968   &7.91    &14.07 &1.75 &4.39 &0.463  &0.295  &0.439 &$b$  &44.6  &FIES\\
59291   &6.34    &13.51 &--5.75 &0.89 &0.510  &0.330  &0.476  &$a$  &46.8  &FIES\\
60678   &8.31   &10.00 &--1.86 &--2.37 &0.494  &  &  &  &5.4  &FIES\\
73600   &9.06   &11.85 &--6.28 &1.60 &0.601  &  &  &  &4.9  &FIES\\
76300   &9.93   &11.59 &0.78 &--6.12 &0.763 &  &  &  &2.5  &FIES\\
89792   &9.00   &11.01 &0.95 &--1.38 &0.589 &  &  &  &18.7  &UVES\\
89825   &9.66   &51.12 &1.63 &1.15 &1.350 &  &  &  &24.6  &FEROS\\
93190   &7.49   &11.62 &0.59 &5.35 &0.443 &0.279  &0.541  &$a$  &8.0  &UVES\\
101137 &8.24   &11.74 &0.60 &--0.62 &0.422 &0.271  &0.451  &$b$  &19.8  &UVES\\
101911 &6.46   &13.44 &0.50 &--1.17 &1.016 &  &  &  &15.8  &FEROS\\
103738 &4.67   &14.24 &0.26 &--1.73 &0.980 &  &  &  &7.8  &FEROS\\
107528 &7.54   &13.14 &0.66 &0.56 &0.401 &0.269  &0.519  &$b$  &32.2  &UVES\\
112584 &9.12   &15.33 &1.65 &--0.34 &0.620 &  &  &  &2.6  &FIES\\
115100 &8.05   &13.83 &0.74 &--4.15 &0.654 &0.408 &0.398 &$b$  &3.8  &UVES\\
\hline
\end{tabular}
\tablefoot{The first to sixth columns give identification, apparent magnitude, parallax, proper motion and colour of $B-V$ which all are from the Hipparcos catalogue \citep{2007A&A...474..653V}. Column 7 and 8 give the colours of $(b-y)$ and $c1$ in Str$\ddot{\rm o}$mgren's photometric system \citep{1983A&AS...54...55O, 1994A&AS..106..257O}. The flags $a$ and $b$ in column 9 indicate that $b-y$ and $c1$ come from \cite{1983A&AS...54...55O}, \cite{1994A&AS..106..257O} respectively. Rotational velocities calculated by measuring Full width at half maximum (FWHM) of atomic and telluric lines are listed in column 10. Column 11 lists the instruments with which the sibling candidates were observed.}
\end{tiny}
\end{center}
\end{table*}


\section{Observations and data reductions}

Observations were carried out with two telescopes for 26 sibling candidates, while spectra of 7 stars were collected from the ESO archive (Table~\ref{canp}). Eighteen of the candidates were observed at the Nordic Optical Telescope (NOT) using the fibre-fed Echelle Spectrograph (FIES) on January 10 -- 12 in 2012. A solar spectrum was also obtained by observing the sky at daytime. The wavelength range of the spectra is 370 -- 730~nm, with a resolution R $\sim$ 67000 and a signal-to-noise ratio (SNR) $>$ 150/pix for most of the spectra. All the spectra were reduced using the FIEStool\footnote{http://www.not.iac.es/instruments/fies/fiestool/FIEStool.html} pipeline. The pipeline includes the following steps to reduce the observed frame: subtracting bias and scattered light, division by a normalized 2-dimensional flat field, extracting individual orders, and finding a wavelength solution and applying it. Finally, all individual spectral orders are merged into a 1-dimensional spectrum. 

Spectra for 12 stars were observed using the UVES spectrograph \citep{2000SPIE.4008..534D} on the VLT 8-m telescope between 2011 and 2012 in service mode. Using image slicer \#3 and with a 0.3$''$ slit width, a resolution  R $\sim$ 110000 was reached in the red arm. The spectra were recorded on three CCDs with wavelength coverages 376--498 nm (blue CCD), 568--750 nm (lower red CCD), and 766--946 nm (upper red CCD). We found that the average SNR of one spectrum in three wavelengths is larger than 160 for all the spectra. The data were reduced with the UVES pipeline. Spectra of 4 of these 12 stars had already been observed with FIES. Although it has been noted that the use of different spectrographs do not introduce significant systematic differences in the derived stellar parameters \citep{2004A&A...415.1153S}, it is still worth to do a further analysis to test for systematic differences (see Sect.~4.2.2).

The spectra for 7 stars extracted from the ESO archive were observed between 2007 and 2010 with FEROS on the ESO 2.2-m telescope \citep{1999Msngr..95....8K}. We checked that the SNR values for these spectra are larger than 100. Since the ESO archive offers reduced 1D spectra with wavelength range of 350 -- 920 nm, we use these spectra for our study. Radial velocities for all spectra were measured by cross-correlation with the solar synthesis spectrum based on the IRAF \footnote{IRAF is distributed by the National Optical Astronomy Observatory, which is operated by the Association of Universities for Research in Astronomy (AURA) under cooperative agreement with the National Science Fundation.} task $\mathit{XCSAO}$. The spectra were also shifted to rest wavelength for radial velocity with the IRAF task $\mathit{DOPCO}$. Their radial velocities are listed in Table~\ref{csp} including their standard deviation.


\section{Spectral analysis}

For our spectral analysis, we use Spectroscopy Made Easy (SME, \citealt{1996A&AS..118..595V, 2005ApJS..159..141V}) to determine the stellar parameters for each star, namely, the effective temperature ($T_{\rm eff}$), surface gravity ($\log g$), and metallicity ([Fe/H]), elemental abundances, and micro-turbulence ($v_{\rm mic}$). SME uses the Levenberg-Marquardt (LM) algorithm to optimize stellar parameters by fitting observed spectra with synthetic spectra. The LM algorithm combines gradient search and linearization methods to determine parameter values that yield a chi-square ($\chi^{2}$) value close to the minimum. Initial stellar parameters ($T_{\rm eff}$, $\log g$, and [Fe/H]) and atomic line data are required to generate a synthetic spectrum. In addition to specified narrow wavelength segments of the observed spectrum, SME requires line masks in order to compare with synthetic spectrum and determine velocity shifts, and continuum masks which are used to normalize the spectral segments. The homogeneous segments and masks are created to fit  all of our solar sibling candidates. 

In SME the model atmospheres are interpolated in the precomputed MARCS model atmosphere grid \citep{2008A&A...486..951G}, which have standard composition. The MARCS grid in SME includes $T_{\rm eff}$ = 2500--8000 K in steps of 100 K from 2500 to 4000 K and 250 K between 4000 and 8000 K, $\log g$ = --0.5 to 5.0 in steps of 0.5, and metallicities between --5.0 to 1.0 in variable steps.

\subsection{The line list}

The elemental abundance derived from a single spectral line is directly proportional to the oscillator strength ($\log gf$) for that line. Therefore, as \cite{2003A&A...410..527B} pointed out, the highest priority is to find homogeneous and accurate $\log gf$ values and we have to make a decision between laboratory and astrophysically determined $\log gf$ values. In addition, the elemental abundance can also be altered by blends. In Table~\ref{loggf}, we list 110 clean iron (80 \ion{Fe}{i} and 20 \ion{Fe}{ii}) lines selected from \cite{2003A&A...410..527B} and Gaia--ESO compiled line list (Heiter et al., in prep). The $\alpha$-elements (Mg, Si, Ca, Ti), iron peak elements (Ni, Cr), sodium and aluminum lines were also selected from those two catalogues. All our clean lines were examined on the Sun spectrum.

\begin{table}[htbp]
\centering
\setlength{\tabcolsep}{11pt}
\begin{tiny}
\caption{Atomic line data.}
\label{loggf}
\begin{tabular}{cccrrr}
      \hline
      \hline
Element    &$\lambda$    &$\chi$    &$\log gf$    &Ref.\\
                   &(${\rm \mathring{A}}$)    &(eV)       \\
      \hline
\ion{Na}{i}    &5688.20    &2.104    &--0.420    &(1)\\
\ion{Na}{i}    &5889.95    &0.000    &0.108       &(2)\\
\ion{Na}{i}    &5895.92    &0.000    &--0.144    &(2)\\
\ion{Na}{i}    &6154.23    &2.102    &--1.510    &(1)\\
.  &.  &.  &.  &. \\
.  &.  &.  &.  &. \\
.  &.  &.  &.  &. \\
      \hline
\end{tabular}
\tablefoot{Columns 1 gives the element with a degree of ionization (\ion{}{i} = neutral, \ion{}{ii} = singly ionized). The wavelength, excitation potential and adopted $\log gf$ values are listed in the columns 2, 3 and 4, respectively. References to the adopted $\log gf$ are given in the column 5, and the corresponding reference is listed at the bottom of the table. The full table is available in electronic form. References:(1) \cite{2003A&A...410..527B}; (2) \cite{1996PhRvL..76.2862V}.}
\end{tiny}
\end{table}

\subsection{The methods to estimate stellar parameters}

Our spectroscopic analysis requires that we estimate six free parameters: $T_{\rm eff}$, $\log g$, [Fe/H], $v_{\rm mic}$, $v_{\rm mac}$, and $v{\rm \sin}i$. We have developed two procedures to efficiently and accurately determine these parameters through multi-step processes.

\subsubsection{Procedure 1: Purely spectroscopic parameters}

In this procedure, initial values for $T_{\rm eff}$, $\log g$, and [Fe/H] are input (the specifics of which are described in Sec. 4.3 and 4.4 for the solar sibling candidates and benchmark stars, respectively). An initial value for $v_{\rm mic}$ was obtained using the relation given in \cite{2013arXiv1309.1099J}, which was derived for stars in the Gaia--ESO survey. Initial values for $v{\rm \sin}i$ were determined by measuring the difference in the full width at half maximum (FWHM) of atomic and telluric lines in the spectrum. This was done using Spectra Visual Editor software (private communication with S. Blanco-Cuaresma). Finally, the initial $v_{\rm mac}$ value was set to 3.0 km~s$^{-1}$ for all stars. Since we are solving for several free parameters simultaneously, there is a degeneracy in the effect of each parameter on the strength of the absorption lines in the spectra. Thus, we have developed a multi-step process to effectively break these degeneracies. The iterative procedure (hereafter procedure 1) follows these steps:
\begin{enumerate}
\item Only $v_{\rm mac}$ is free while other parameters are fixed;\\
\item Only $v{\rm \sin}i$ is free and fixed $v_{\rm mac}$ comes from step 1;\\
\item Only [Fe/H] is free and fixed $v{\rm \sin}i$ is from step 2;\\
\item Both $v_{\rm mic}$ and $v{\rm \sin}i$ are free and fixed [Fe/H] comes from step 3;\\
\item Repeat the steps 1 to 4 with updated parameters ($v_{\rm mic}$ and $v{\rm \sin}i$) until $v_{\rm mac}$, $v_{\rm mic}$ and $v{\rm \sin}i$ converge;\\
\item $T_{\rm eff}$ $\log g$, and [Fe/H] are free and the other fixed parameters are from step 5;\\
\item Repeat the steps 1 to 6 with updated parameters ($T_{\rm eff}$, $\log g$, and [Fe/H]) until all 6 parameters reach convergence.
\end{enumerate}

\subsubsection{Procedure 2: Using parallax estimates}

It has been shown that analysis, which compute stellar parameters using purely spectroscopic means, can lead to erroneous results. For example, \cite{2014A&A...562A..71B} identified that for dwarf stars with $\log g$ > 4.2 enforcing ionization equilibrium between \ion{Fe}{I} and \ion{Fe}{II} lines does not yield accurate $\log g$ estimates. Our analysis could also suffer from such systematic effects. Thus, we have also developed a second procedure (hereafter procedure 2) in which we determine $\log g$ using parallax estimates for the stars.

We compute $\log g$ using the following equation:
\begin{equation}
\log g=4\log T_{\rm eff} + 0.4V -2\log(1/\varpi) - 0.4\rm B.C.
            + \log(M/M_{\odot}) - 10.5037,
\label{sg}
\end{equation}
where $V$, $\varpi$, M and B.C. are the apparent magnitude, parallax, stellar mass and bolometric correction, respectively. Since the stars are located in the Local Bubble \citep{2003A&A...411..447L}, we assume that the extinction is negligible (comparing with \citealt{2014A&A...568A..25N}). \cite{1996ApJ...469..355F} expressed the bolometric corrections as a function of $T_{\rm eff}$ and found that all luminosity classes appear to follow a unique $T_{\rm eff}-\rm B.C.$ relation. We thus used this relation, utilizing the corrected coefficients from \cite{2010AJ....140.1158T} to obtain B.C. for our stars. The stellar mass and age of each star require to fit isochrones using our estimated values for $T_{\rm eff}$, $\log g$, and [Fe/H]. Thus, we must iterate until the input stellar parameters and output mass and age converge.

Procedure 2 is defined by the following approach. Firstly, using procedure 1, $T_{\rm eff}$ and [Fe/H] were computed by fixing $\log g$ in SME. Then, the mass and age were obtained through fits to the Yonsei-Yale isochrones \citep{2004ApJS..155..667D}, by maximising the probability distribution functions, as described in \cite{2011A&A...533A.134B}. Subtituting these values of $T_{\rm eff}$ and mass into equation \eqref{sg}, new $\log g$ values were calculated. We then returned to the first step in procedure 1 and recomputed $T_{\rm eff}$ and [Fe/H], holding $\log g$ fixed. These new values were then used to fit to the isochrones to get new estimates of mass and age. We iterated until convergence between the $\log g$, mass and age estimates. The final values for $T_{\rm eff}$, $\log g$ and [Fe/H] were obtained when the stellar parameters converge and the average differences of stellar ages and masses from two iterations are less than 0.1 Gy and 0.01 solar mass, respectively.

\subsection{Stellar parameters for solar sibling candidates}

\subsubsection{Initial stellar parameters}

Initial effective temperatures for the stars were determined using both Str$\ddot{\rm o}$mgren ($uvby$) and $UBV$ photometry \citep{1983A&AS...54...55O, 1994A&AS..106..257O} (Table~\ref{canp}). Str$\ddot{\rm o}$mgren photometric system is specially designed to measure the physical properties of the stellar atmospheres and the colour $(b-y)$ is very sensitive to the effective temperature. The calibration of $T_{\rm eff}$ versus $(b-y)-c_{1}-\rm [Fe/H]$ from \cite{1996A&A...313..873A} (their Eq. 9) was used. The $uvby$ data is taken from \cite{1983A&AS...54...55O, 1984A&AS...57..443O, 1994A&AS..106..257O}. However, some of our candidates were not included in those catalogues. Since the Hipparcos catalogue offers $B-V$ for all of our stars, the relationship of $(B-V) - T_{\rm eff}-\rm [Fe/H]$ from \cite{1996A&A...313..873A} (their Eq. 2) was also used to calculate $T_{\rm eff}$. The standard deviation for this calibration of $T_{\rm eff}$ is 130 K, which implies a precision of 2.2\% at the Sun's temperature (5777 K). For the Str$\ddot{\rm o}$mgren, $T_{\rm eff}$ the standard deviation is 110 K.

\cite{1996A&A...313..873A} found that an error of 0.3 dex in [Fe/H] implies a mean error of 1.3\% in $T_{\rm eff}$. For our solar sibling candidates, it is safe to assume that all of them have solar metallicity. This is supported by the small variation of [Fe/H] obtained from our abundance analysis. The average of the two effective temperatures was used as our initial guess for $T_{\rm eff}$.

Initial gravities were determined using Eq. \eqref{sg} and the photometric estimated of $T_{\rm eff}$. We further assumed that all the candidates have solar metallicity and age ($\sim$4.6 Gyr). As the effective temperatures and absolute magnitudes ($M_{\rm V}$) of candidates are obtained in the previous, we can obtain the masses for all candidates by interpolating $T_{\rm eff}$ and $M_{\rm V}$ within  isochrones Yonsei-Yale isochrones \citep{2004ApJS..155..667D}.

\subsubsection{Best stellar parameters}	

\begin{figure*}[htbp]
\begin{center}
\includegraphics[scale=0.45]{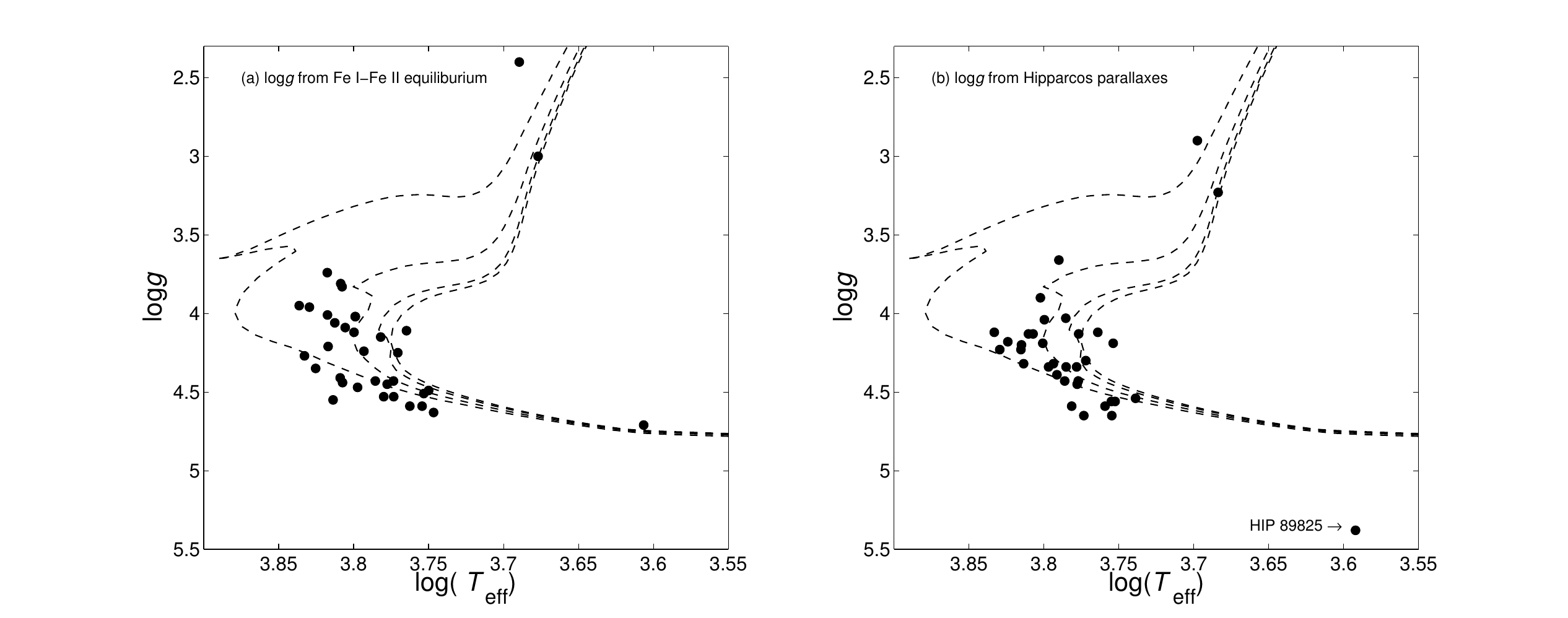}
\caption{(a) HR diagram for the sample when (a) log$g$ is based on \ion{Fe}{I}-\ion{Fe}{II} ionization equilibrium, and (b) when log$g$ is based on Hipparcos parallaxes.  Four isochrons at solar metallicity and four different ages (1, 3, 5, and 6 Gyr) according the the Yonsei-Yale models \citep{2004ApJS..155..667D} are also shown.}
\label{comcmd}
\end{center}
\end{figure*}

All stellar parameters derived using procedures 1 and 2 are given in Table~\ref{csp}. In Fig.~\ref{comcmd}, we show a comparison of the result that stellar parameters from each procedure in the H-R diagram. As can be seen in Fig.~\ref{comcmd}a, some stars with $\log g$ > 4.2 appear to fall in regions un-occupied by the isochrones, when stellar parameters were derived using purely spectroscopic means. On the other hand, when we used $\log g$ derived using the parallax, we see improvements in each star's location in the H-R diagram (Fig.~\ref{comcmd}b). This is in part due to the fact that we use the isochrones to derive our parallax gravities. Even though the isochrones have their own associated uncertainties (corresponding to uncertainties in stellar evolution theory), we expect the systematic effect on the derived parallax gravities to be less than 0.15~dex (see Eq. \eqref{sg}). Thus, for the remainder of our analysis we adopt the stellar parameters derived using procedure 2.

In Fig. \ref{comcmd}b it should be noticed that one star --HIP 89825-- has unexpected large $\log g$ and is far below the isochrones. The reason could be that we got the wrong bolometric correction for this star. As \cite{2010AJ....140.1158T} pointed out that bolometric corrections become less reliable for cooler stars and break down completely for M dwarfs, the obtained bolometric correction of HIP 89825, which is a typical M dwarf, could be far away the real one. It also should be mentioned that assumptions of solar metallicity and age were made in order to derive the initial $T_{\rm eff}$ and stellar mass. However, HIP 89825 has much lower metallicity than that of the Sun. All suggest that a unreliable $\log g$ from the parallax was estimated. However, $\log g$ from pure spectroscopic approach has reasonable value shown in Fig. \ref{comcmd}a. In this case, the stellar parameters from pure spectroscopic approach were used to determine the elemental abundances for this star.

Although our iron line list was slightly different for the FIES and UVES spectra because of different wavelength observations, we found that four of five stars, including the Sun, observed by both the FIES and UVES instruments have quite similar outputs in stellar parameters. The mean differences of stellar parameters are within the estimated uncertainties. This suggests that the our analysis give the same results independent of the
spectra and their resolutions. Outputs of $v_{mic}$, $v_{mac}$, and $v{\rm \sin}i$ for two spectra of a star, HIP 9405, are not far from each other. However, they have total different values in $T_{\rm eff}$, $\log g$, and [Fe/H]. Comparing two spectra, it clearly shows that the two spectra come from two stars. Recently a study concluded that HIP 9405 belongs to a binary system \citep{2007A&A...464..377F}. It is highly possible that two spectra come from two companion stars, respectively. We can not identify which spectrum comes from our sibling candidate. Thus, we exclude this star from the remainder of our analysis. We carefully inspected all spectra and found that the spectra of HIP 56798 and HIP 25358 have clear double line signature. They could be companion stars of binaries. It could bring larger uncertainties than our given typical errors on stellar parameters caused by near-by line blending. It also should be noticed that the stellar parameters of fast rotational stars ($v{\rm \sin}i$ > 30 km s$^{-1}$) might suffer larger uncertainties than the typical errors.

\begin{table*}[htbp]
\centering
\setlength{\tabcolsep}{5pt}
\begin{tiny}
\caption{Stellar parameters of sibling candidates.}
\label{csp}
\begin{tabular}{cccrrrrcccrrrrr}
      \hline
      \hline
      Names  &$T_{\rm eff}$  &$\log g$  &[Fe/H]  &$v_\mathrm{mic}$ &$v_\mathrm{mac}$ &$v{\rm \sin}i$ &$T'_{\rm eff}$  &log $g'$  &[Fe/H]$'$  &$v_\mathrm{mic}'$ &$v_\mathrm{mac}'$ &$v{\rm \sin}i'$&V$_{\rm r}$&$\sigma(\rm V_{r})$\\
        HIP          &(K)     &($\mathrm{cm~s}^{-2}$)    &    &($\mathrm{km~s}^{-1}$)     &($\mathrm{km~s}^{-1}$) &($\mathrm{km~s}^{-1}$) &(K)     &($\mathrm{cm~s}^{-2}$)    &    &($\mathrm{km~s}^{-1}$)     &($\mathrm{km~s}^{-1}$) &($\mathrm{km~s}^{-1}$)&($\mathrm{km~s}^{-1}$)&($\mathrm{km~s}^{-1}$)\\
      \hline
      7764             &6046  &4.59  &0.11    &1.0  &4.0  &1.7   &5992  &4.45  &0.14    &1.0  &3.9  &1.8  &--24.6  &0.1\\
      8444             &6111  &4.43  &--0.10    &1.0  &3.7  &6.2   &6271  &4.47  &0.03    &1.2  &3.4  &6.3  &0.5  &0.2\\
    10786             &5995  &4.45  &--0.07   &1.0  &6.6  &6.9   &6102  &4.43  &0.03   &1.2  &6.6  &6.8  &--13.6  &0.2\\
    14640$^{a}$ &6461 &4.13  &--0.09  &1.5  &5.8  &14.6 &6494 &4.06  &--0.01  &1.6  &5.2  &14.9  &23.6  &0.2\\
    14640$^{b}$ &6415 &4.13  &--0.14     &1.5  &5.3  &14.6 &6568 &4.01  &0.02     &1.6  &5.9  &14.6  &23.6  &0.2\\
     15929            &6184 &4.39  &--0.19     &1.2  &11.4  &36.4 &6512 &4.55  &0.09     &1.3  &2.9  &37.8   &13.1  &0.3\\
     21158            &5916 &4.30  &0.09     &1.1  &3.2  &1.7   &5895 &4.25  &0.11     &1.1  &2.6  &2.6  &6.6  &0.1\\
     22002            &5481 &4.54  &0.16     &1.2  &1.7  &2.9   &5580 &4.63  &0.28     &1.2  &0.3  &3.2  &13.1  &0.2\\
     24232            &6531 &4.20  &0.03     &1.5  &8.5  &13.4 &6561 &4.21  &0.10     &1.5  &6.3  &14.1  &12.2  &0.3\\
     25358            &6526 &4.23  &0.24     &1.2  &8.4  &14.4 &6305 &4.12  &0.14     &1.2  &5.8  &15.0  &39.7  &0.4\\
     26744            &6344 &3.90  &--0.05     &1.7  &6.8  &8.9   &6436 &3.81  &0.05     &1.7  &6.4  &9.2  &--5.1  &0.3\\
     30344            &5744 &4.59  &0.07     &1.0  &3.2  &2.6   &5787 &4.59  &0.12     &1.1  &3.2  &2.8  &29.2  &0.1\\
     33275            &6218 &4.32  &--0.09  &1.2   &6.3  &5.3  &6210 &4.24  &--0.07  &1.3   &3.7  &7.0  &--14.3  &0.2\\
     33685            &6752 &4.23  &--0.06     &1.8   &--0.3 &73.4 &6804 &4.27  &0.03     &1.8   &--0.3 &73.4  &27.3  &1.1\\
     40317$^{a}$ &5688 &4.56  &0.05    &0.8   &2.9  &0.7  &5664 &4.51  &0.03    &1.0   &3.0  &0.8  &34.2  &0.1\\
     40317$^{b}$ &5656 &4.56  &0.03    &0.8   &3.5  &1.0  &5622 &4.49  &0.02    &1.0   &3.4  &1.0  &34.2  &0.1\\
     48062$^{a}$ &6264 &4.34  &--0.16    &1.2   &8.5  &19.6 &6418 &4.44  &0.00    &1.2   &8.3  &19.7  &3.0  &0.3\\
     48062$^{b}$ &6098 &4.34  &--0.26    &1.0   &5.6  &19.6 &6441 &4.41  &0.03    &1.2   &10.3  &19.6  &3.0  &0.3\\
     51581             &5981 &4.13  &--0.05    &1.3   &3.8  &5.8   &6052 &4.15  &0.10    &1.3   &3.1  &6.1  &16.5  &0.1\\
     53921             &6305 &4.04  &--0.14  &1.5  &3.5  &5.5   &6293 &4.02  &--0.06  &1.5  &4.9  &4.6  &17.4  &0.1\\
     56798             &5672 &4.19  &0.00     &0.5  &4.2  &2.6   &5934 &4.43  &0.17     &0.9  &5.4  &0.5  &11.9  &0.2\\
     58968             &6509 &4.32  &0.01     &1.4  &12.1 &30.6 &6688 &4.35  &0.17     &1.5  &7.9 &31.5  &--14.3  &0.3\\
     59291             &6166 &3.66  &--0.10     &1.7  &7.3   &31.8 &6422 &3.82  &0.14     &1.7  &6.0   &31.9  &--20.6  &0.3\\
     60678             &6101 &4.03  &--0.24  &1.2   &7.5   &3.3   &6294 &4.02  &--0.22  &1.4   &3.2   &6.9  &--27.6  &0.3\\
     73600             &5985 &4.43  &0.10     &1.3   &5.2   &5.1   &6024 &4.53  &0.18     &1.3   &4.8   &5.4  &3.7  &0.1\\
     76300             &5685 &4.65  &0.18     &0.8   &3.4   &1.5   &5676 &4.60  &0.17     &1.0   &3.3   &1.2  &--11.4  &0.1\\
     89792             &6000 &4.34  &--0.11     &0.5   &8.1   &42.6 &6391 &4.09  &0.33     &1.0   &12.4   &45.9  &6.6  &0.6\\
     89825             &3925 &5.37  &--0.03   &0.1  &0.1  &0.15  &4041 &4.71  &--0.38  &0.5   &1.3   &0.2  &--38.8  &1.0\\
     93190             &6809 &4.12  &0.14     &1.9   &6.4   &7.9    &6752 &3.96  &0.17     &1.9   &5.6   &8.4  &--31.3  &0.3\\
   101137             &6321 &4.19  &--0.35  &1.1   &3.6   &33.4 &6570 &3.74  &--0.11  &1.3   &3.0   &34.7  &--5.5  &0.4\\
   101911             &4828 &3.23  &--0.02     &1.2  &2.4   &1.9   &4752 &3.00  &0.02     &1.1  &2.4   &1.5  &26.6  &0.5\\
   103738             &4984 &2.90  &--0.02   &1.0  &4.8   &6.8   &4892 &2.40  &--0.03   &1.1  &5.4   &6.6  &18.0  &0.2\\
   107528             &6669 &4.18  &0.01      &1.8  &6.0   &18.1  &6859 &3.95  &0.13      &2.0  &3.7   &18.3  &--6.4  &0.3\\
   112584             &5934 &4.65  &--0.04   &0.8  &5.2   &0.8     &5932 &4.53  &--0.03   &1.1  &5.5   &0.3  &--0.3  &0.1\\
   115100             &5808 &4.12  &0.08      &1.2  &2.4   &3.8     &5817 &4.11  &0.18      &1.2  &2.2   &4.0  &--24.3  &0.1\\
           \hline
\end{tabular}
\tablefoot{The first column gives identification of each stars. The second to seventh columns give the global parameters obtained through our methodology (Procedure 2) that surface gravity is calculated from parallax and temperature and effective temperature, metallicity, micro-turbulence, marco-turbulence, and rotation velocity are computed by fitting spectra. The eighth to thirteenth columns give the six parameters obtained from the purely spectroscopic approach (Procedure 1). The unprimed quantities are the preferred values according to Sect. 4.3.2. $^{a}$ indicates that the parameters are measured based on the FIES's spectrum, while $^{b}$ means that the parameters are measured based on the UVES's spectrum.}
\end{tiny}
\end{table*}

\subsection{Estimating systematic uncertainties}

In order to determine the systematic errors in our derived stellar parameters from procedure 2, we applied our analysis to several standard stars for which accurate stellar parameters have been estimated by other means.

\subsubsection{Gaia Benchmark stars}

Recently, a set of reference stars have been created for calibration purposes in the Gaia mission. For these benchmark stars $T_{\rm eff}$ and $\log g$ are well determined independently from spectroscopy. Effective temperatures of benchmark stars are directly determined from angular diameters and bolometric fluxes. Surface gravities are also directly measured from the stellar mass and radius which are calculated based on angular diameter and parallax, while the metallicity of each benchmark star has been derived through spectroscopic \citep{2013arXiv1312.2943J}.

In the current work, the Sun and four benchmark stars (listed in Table~\ref{bm}) are used to test our methods, derive systematic errors and fix the linelist. These benchmark stars are very similar to the Sun in metallicity ($\Delta {\rm [Fe/H]} < \pm 0.1$) and cover the same $T_{\rm eff}$ and $\log g$ range as our sibling candidates. Ten high SNR spectra of four benchmark stars and the Sun collected from the UVES archives \citep{2000SPIE.4008..534D}, NARVAL\footnote{http://www.ast.obs-mip.fr/projets/narval/v1/}, HARPS \citep{2003Msngr.114...20M} and UVES-POP library \citep{2003Msngr.114...10B} were analysed. 

\begin{table*}[htbp]
\begin{center}
\setlength{\tabcolsep}{5pt}
\begin{tiny}
\caption{Stellar parameters for the Sun and four benchmark stars used to develop our methodology.}
\label{bm}
\begin{tabular}{lrrr|rcrcc|ccrc}
      \hline
      \hline
              &        &         &   &\multicolumn{5}{c}{Recommended}     &\multicolumn{4}{|c}{This study}\\
      ID1 & ID2 & RA &DEC &$T_{\rm eff, r}$ &$\log g_{\rm r}$  &[Fe/H]$_{\rm r}$ &$v{\rm \sin}i_{\rm r}$ &Ref.  &$T_{\rm eff, p}$ &$\log g_{\rm p}$ &[Fe/H]$_{\rm p}$ &$v{\sin}i_{\rm p}$\\
                 &     &("hh:mm:ss")&("dd:mm:ss")     &(K)    & ($\mathrm{cm~s}^{-2}$)  &        &($\mathrm{km~s}^{-1}$)   &      &(K)    &($\mathrm{cm~s}^{-2}$)   &      &($\mathrm{km~s}^{-1}$)\\
      \hline
      Sun        &                  &                         &                         &5777 &4.44 &0.00   &1.6 &m   &5793 &4.44 &0.06 &1.7\\
$\beta$~Hyi &HIP 2021 &00 25 45.070 &--77 15 15.29 &5873 &3.98 &--0.09 &3.3 &a, n, b, r, p/yy &5890 &4.03 &--0.02 &1.5 \\
$\epsilon$~Eri &HIP 16537 &03 32 55.845 &--09 27 29.73 &5050 &4.60 &--0.07  &2.4 &a, d, c, v, p/yy  &5143 &4.69 &--0.05 &2.4\\
Procyon &HIP 37279 &07 39 18.119 &+05 13 29.96 &6545 &3.99 &0.00 &2.8 &a, e, g, p/yy  &6640 &4.07 &0.02 &4.7\\
18~Sco &HIP 79672 &16 15 37.269 &--08 22 09.99 &5747 &4.43 &0.02 &2.2 &a, k, h, s, p/yy  &5878 &4.52 &0.13 &2.0\\
	\hline
\end{tabular}
\tablefoot{The first to fourth columns give two identifications and equatorial coordinates. The recommended values of metallicity, effective temperature, surface gravity, and rotation velocity are listed in columns 5 to 8, while our results are shown in column 10 to 13. The recommended metallicities of benchmark stars are obtained from \cite{2013arXiv1312.2943J} which is indicated in a, while the recommended $T_{\rm eff, r}$ and $\log g_{\rm r}$ are directly determined from angular diameter, bolometric flux, parallax and stellar mass. Flag m indicates reference \cite{2012MNRAS.422..542P}. Flags n, d, e, and k indicate that the data of angular diameters come from references \cite{2007MNRAS.380L..80N}, \cite{2004A&A...426..601D}, \cite{2005ApJ...633..424A}, and \cite{2004A&A...426..297K}. Flags b, c, and h indicate that the data of bolometric fluxes are collected from \cite{1998A&AS..129..505B},  \cite{1996A&AS..117..227A}, and \cite{1995A&A...297..197A}. Flags r, v, g, and s indicate that rotation velocities are selected from \cite{2003A&A...398..647R}, \cite{2005ApJS..159..141V}, \cite{2010MNRAS.405.1907B}, and \cite{1997MNRAS.284..803S}. p/yy means that average of two masses which are from Padova \citep{2008A&A...484..815B} and Yonsei-Yale (\cite{2003ApJS..144..259Y}, \cite{2004ApJS..155..667D}) stellar evolutionary tracks using direct $T_{\rm eff}$ and luminosity.
}
\end{tiny}
\end{center}
\end{table*}

\subsubsection{Estimating systematic errors in our stellar parameters}

\begin{figure}[htbp]
\begin{center}
\includegraphics[width=9.5cm, height=7cm]{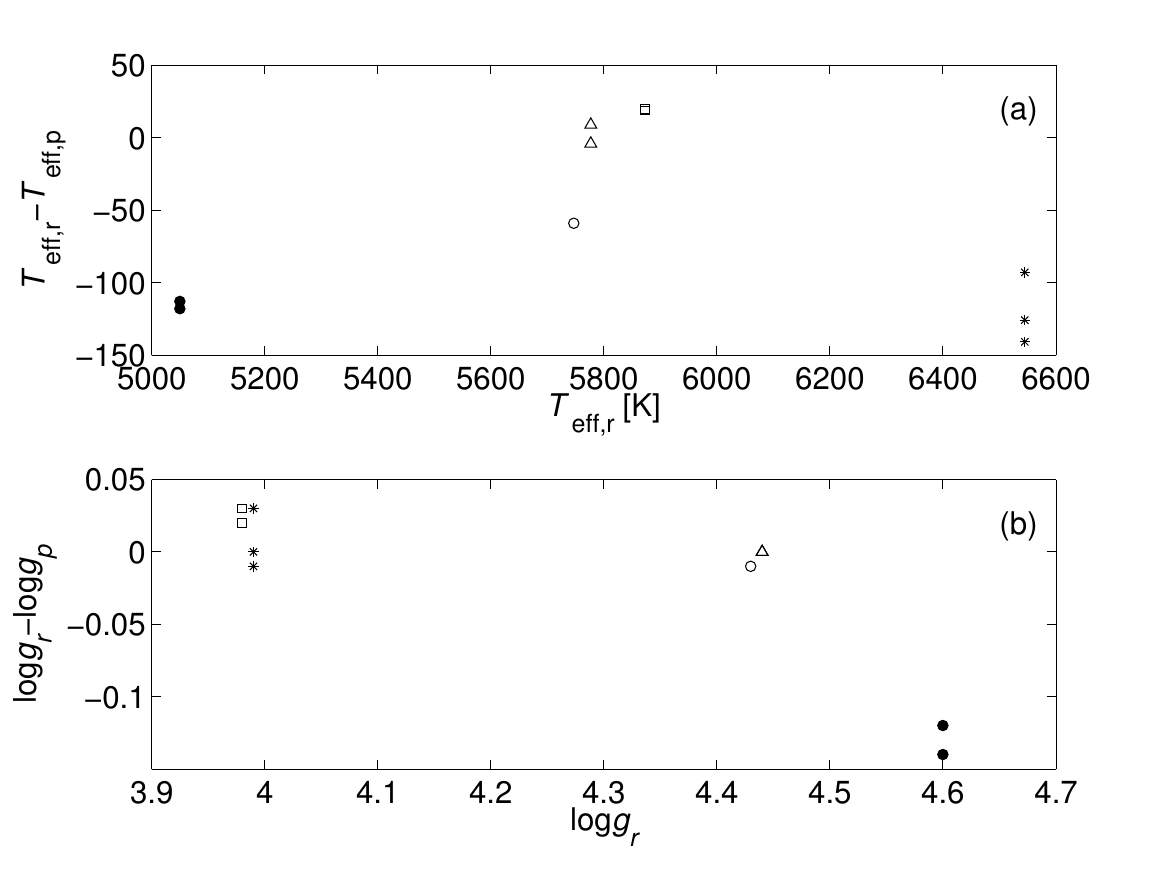}
\caption{(a) $T_{\rm eff,r}-T_{\rm eff,p}$ vs. $T_{\rm eff,p}$, where $T_{\rm eff,r}$ represents the recommended effective temperature, and $T_{\rm eff,p}$ is the best determined effective temperature from our methodologys. Five symbols stand for the five benchmark stars, while two (or three) of the same symbol indicates that spectra obtained with different instruments were analysed. It is the same for the log $g_{\rm r}$ and log $g_{\rm p}$ which are shown in panel (b). }
\label{bmtg}
\end{center}
\end{figure}

Although $T_{\rm eff}$ and $\log g$ are well determined for the benchmark stars, we recalculated them using photometry and astrometry data to emulate our exact methodology for our solar sibling candidates. Fig. \ref{bmtg} shows the difference between the recommended values listed in Table~\ref{bm} (given by $T_{\rm eff,r}$ and log $g_{\rm r}$) and those derived by our methods. It was found that the mean difference in $T_{\rm eff}$, $\log g$, [Fe/H] and $v{\rm \sin}i$ and their standard deviation are --67$\pm$40 K, --0.08$\pm$0.06 dex, --0.05$\pm$0.03 dex, and 0.2$\pm$1.3 $\mathrm{km~s}^{-1}$, respectively. It should be noticed that the typical uncertainties of the recommended $T_{\rm eff}$ and $\log g$ are about 50 K and 0.02 dex for the benchmark stars, respectively. The random errors of $\log g$ obtained from distances and temperatures are between 0.04 to 0.06 dex. It is consistent with the scatter of mean difference of $\log g$. Considering these systematic errors and possible sources of uncertainty on atmospheric model and atomic line data, the systematic errors in the stellar parameters were estimated to be $\delta T_{\rm eff}=67$ K, $\delta {\rm log}$ $g = 0.08$ dex and $\delta v{\rm \sin}i = 0.2$ $\mathrm{km~s}^{-1}$.

\begin{figure}[htbp]
\begin{center}
\includegraphics[width=9.5cm, height=7cm]{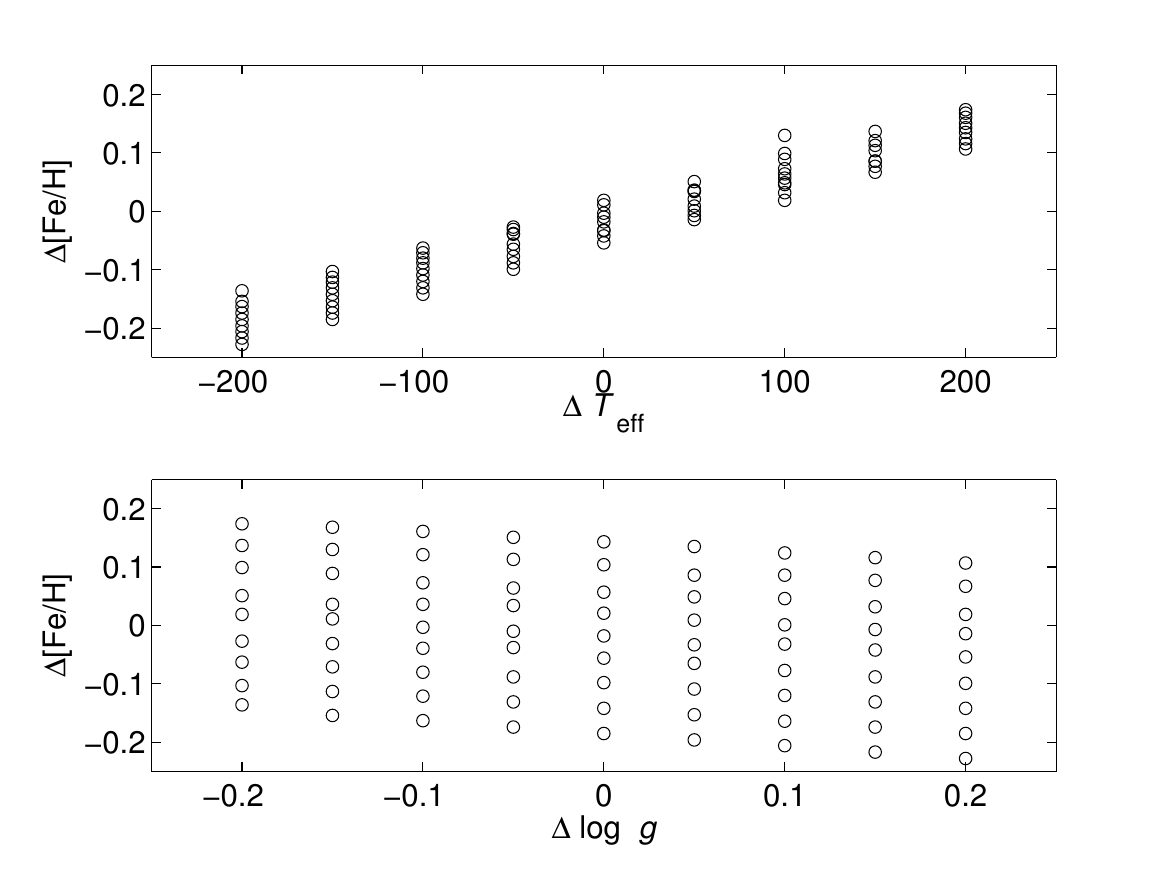}
\caption{Sensitivity of [Fe/H] to changes in stellar parameters for the Sun. A significant correlation between $\Delta$[Fe/H] and $\Delta T_{\rm eff}$ exists, while the trend between $\Delta$[Fe/H] and $\Delta$$\log g$ is hardly detected.}
\label{bmm}
\end{center}
\end{figure}

Although the recommended metallicities were determined from high resolution spectra, it might suffer a large uncertainty. It is known that [Fe/H] depends on both $T_{\rm eff}$ and $\log g$. We can simulate the distribution of errors of [Fe/H] by iterating different $T_{\rm eff}$ and $\log g$ values. A two dimensional grid was created with varying $T_{\rm eff}$ and $\log g$. Since we have estimated errors of $T_{\rm eff}$ and $\log g$ from our benchmark stars analysis, we assumed that initial values of $T_{\rm eff}$ and $\log g$ vary within 3$\sigma$ in steps of 50 K and 0.05, respectively. The spectra of the Sun, $\epsilon$Eri, and HIP 51581 were used to calculate $\Delta$[Fe/H], the difference between the newly output [Fe/H] and those obtained from procedure 2. Fig. \ref{bmm} shows how $\Delta$[Fe/H] significantly correlates with $\Delta T_{\rm eff}$ for the one of three stars. On the other hand, the trend between $\Delta$[Fe/H] and $\Delta$$\log g$ is hardly detected, because $\log g$ is sensitive to Fe II rather than Fe I and most of the selected lines are Fe I in our line list.

As no significant correlation between $\Delta$[Fe/H] and $\Delta$$\log g$ was found,  the change of [Fe/H] responding to the error of $\log g$ is much smaller than that with $\Delta T_{\rm eff}$. It was also found that the maximum uncertainty of [Fe/H] caused by the typical error in $v{\rm \sin}i$ is 0.01 dex within our three tested stars. The change in [Fe/H] is still very small between 0.01 -- 0.02 dex, if we vary the $v_{\rm mic}$ by 0.1 $\mathrm{km~s}^{-1}$. We then could simply use the root sum square of all changes to calculate the total uncertainty in [Fe/H]. Finally, uncertainties of [Fe/H] are between 0.04 to 0.06 dex for spectra with different SNR. It is consistent with the mean difference of [Fe/H] between recommended and our studies for benchmark stars. The maximum uncertainty of 0.06 dex will be regarded as the systematic error of measurement in metallicity.

\subsection{Ages}

The age of each star was determined during the fits to the isochrones in procedure 2, as described in Sec. 4.2.2. The most probable age is determined from the peak of the age probability distribution, 1$\sigma$ lower and upper age limits are obtained from the shape of the distribution. Stellar masses were also determined in a similar manner. Both ages and masses are reported in Table~\ref{mage}. It clearly shows that most of solar sibling candidates have younger ages than that of the Sun in Fig. \ref{comcmd}. 

It is possible that the probabilistic age determinations used in our analysis suffer from systematic biases mainly caused by sampling the isochrone data points \citep{2004A&A...418..989N}. Age degeneracy around the zero-age main sequence or at the turn-off could also induce systematic effects because of complex isochrones. The Bayesian approach proposed by \cite{2005A&A...436..127J} was used to cope with these problems. In order to find out if possible systematic biases exist in our probabilistic determinations, we used the \cite{2006A&A...458..609D} PARAM web interface\footnote{http://stev.oapd.inaf.it/cgi-bin/param} bayesian based method to estimate the stellar ages. Comparing our ages with the PARAM determined ages, the average difference is --0.1 Gyr, with a standard deviation 0.6 Gyr, which is much smaller than the typical uncertainties which are of the order 1 Gyr. This suggests that the determined ages of our sample are reliable.

\begin{table}[htbp]
\centering
\setlength{\tabcolsep}{8pt}
\begin{tiny}
\caption{Stellar masses, absolute magnitudes, and ages of solar sibling candidates.}
\label{mage}
\begin{tabular}{cccrrrrcccrrr}
      \hline
      \hline
      Names  &M &$M_{V}$ &$\sigma_{M_{V}}$ &Age &-1$\sigma$ &+1$\sigma$\\
        HIP          &(M$_{\odot}$)&(mag) &(mag) &(Gyr) &(Gyr) &(Gyr)\\
      \hline
      7764              &1.11 &4.79 &0.12 &0.3 &-- &1.5\\
      8444             &1.10 &4.35 &0.17 &2.4 &1.2 &3.8\\
    10786             &1.07 &4.49 &0.12 &3.3 &1.4 &4.6\\
    14640$^{a}$ &1.32 &3.20 &0.12 &2.6 &2.3 &2.9\\
    14640$^{b}$ &1.30 &3.20 &0.12 &3.0 &2.5 &3.3\\
     15929            &1.09 &4.13 &0.15 &3.2 &1.4 &3.9\\
     21158            &1.08 &4.14 &0.05 &5.2 &4.0 &6.4\\
     22002            &0.95 &5.14 &0.18 &4.9 &2.3 &8.7\\
     24232            &1.33 &3.32 &0.12 &2.2 &1.5 &2.4\\
     25358            &1.38 &3.39 &0.14 &1.3 &0.8 &1.7\\
     26744            &1.41 &2.59 &0.08 &3.0 &2.3 &3.4\\
     30344            &1.01 &5.03 &0.04 &0.3 &0.6 &2.9\\
     33275            &1.14 &3.95  &0.09 &3.1 &2.0 &3.5\\
     33685            &1.36 &3.22 &0.09 &1.2 &0.9 &1.8\\
     40317$^{a}$ &0.98 &5.05 &0.20 &2.8 &1.6 &6.3\\
     40317$^{b}$ &0.97 &5.05 &0.20 &3.7 &1.7 &6.7\\
     48062$^{a}$ &1.14 &4.04 &0.19 &3.0 &1.4 &3.5\\
     48062$^{b}$ &1.04 &4.04 &0.19 &5.0 &2.9 &6.0\\
     51581             &1.12 &3.65 &0.16 &5.6 &4.7 &6.7\\
     53921             &1.26  &3.13 &0.07 &3.8 &3.0 &4.2\\
     56798             &0.98  &4.13 &0.18 &10.0 &8.5 &11.4\\
     58968             &1.29 &3.65 &0.09 &1.4 &0.7 &1.9\\
     59291             &1.57  &1.99 &0.05 &2.5 &2.1 &2.6\\
     60678             &1.14  &3.31 &0.15 &5.5 &4.6 &6.1\\
     73600             &1.10  &4.43 &0.11 &2.1 &1.2 &3.7\\
     76300             &1.03  &5.25 &0.17 &0.6 &0.8 &3.5\\
     89792             &1.06  &4.21 &0.19 &4.8 &3.2 &6.1\\
     89825             &0.54  &8.20 &0.07 &13.1 &4.5 &13.1\\
     93190             &1.51  &2.82 &0.11 &1.2 &1.1 &1.7\\
   101137             &1.13  &3.59 &0.11 &4.5 &3.5 &4.8\\
   101911             &1.06  &2.10 &0.08  &6.4 &4.6 &11.7\\
   103738             &2.45  &0.44 &0.04  &0.6 &0.6 &1.2\\
   107528             &1.36  &3.13 &0.11 &1.7 &1.3  &2.0\\
   112584             &1.04  &5.05 &0.10 &0.3 &0.3 &1.4\\
   115100             &1.09  &3.75 &0.12  &7.0 &6.2 &8.4\\
           \hline
\end{tabular}
\tablefoot{$^{a}$ and $^{b}$ have the same meanings as in Table~\ref{csp}. "--" indicates the 1$\sigma$ lower (or upper) age could not be determined because very young (or old) star is out of isochrones limitations.}
\end{tiny}
\end{table}

\subsection{Abundances analysis}

We obtained abundances by fitting the selected absorption lines for each element. During the abundance analysis, we left corresponding elemental abundance (e.g. [Na/H]) free while the stellar parameters were kept fixed. The average ratio $A_{\rm{Na}}$ (the absolute abundance relative to the total number density of atoms) was output from SME. The solar elemental abundance pattern taken from \cite{2007SSRv..130..105G} was used as a template for stellar abundances, in order to obtain the abundance ratio [Na/H]. The abundance ratios with respect to Fe (i.e. [X/Fe] in standard notation) were also calculated and are shown in Table~\ref{abun}. We measured abundances for Na, Mg, Al, Si, Ca, Ti, Cr, and Ni using either neutral or both neutral and singly ionized lines, as listed in Table~\ref{loggf}. For comparison, we also derived solar abundances using the same line list, and the stellar parameters derived from our solar spectrum in Sect. 4.3.2. For the solar sibling candidates observed by FIES and FEROS, we determined the elemental abundances (see Table~\ref{abun}) relative to solar values using the spectrum of sky at daytime. For the spectra of targets observed with UVES, the elemental abundances relative to solar values were determined using the solar spectrum reflected of the Moon. Some studies pointed out that systematic biases in solar abundance analysis could be introduced by aerosol and Rayleigh-Brillouin scattering filling up the day sky solar spectrum \citep{2000PASP..112..328G} and using different spectrographs \citep{2014ApJ...795...23B}. In order to find out if possible biases exist in our solar abundances, we measured equivalent widths (EWs) of two solar spectra from both FIES and UVES for all iron lines and found that the average difference of two solar EWs is --1.9$\pm$4.6~m$\AA$. Comparing with typical uncertainties in elemental abundances (see Sect. 4.7), we ignored the systematic error caused by the average difference of EWs.

\begin{table*}[htbp]
\centering
\setlength{\tabcolsep}{12pt}
\begin{tiny}
\caption{The elemental abundances of solar sibling candidates.}
\label{abun}
\begin{tabular}{crrrrrrrrrrrrrc}
      \hline
      \hline
                      &        &---   &---  &---  &---   &[X/Fe]    &---   &---    &---   \\
      HIP  &[Mg/H] &Na &Mg &Al &Si &Ca &Ti &Cr &Ni &$\delta_{C}$ &$P_{\delta_{C}}$\\
               &             &      &       &    &      &     &      &     &     &                   &(per cent) \\
      \hline
      7764             &0.07  &--0.06  &--0.05  &--0.08  &--0.02  &--0.03  &0.02  &0.07  &--0.01  &0.047  &81\\
      8444             &--0.10  &--0.10  &0.00  &--0.15  &0.04  &0.05  &--0.04  &0.05  &--0.08  &0.066  &54\\
    10786             &--0.12  &--0.13  &--0.07  &--0.14  &--0.03  &0.06  &0.00  &0.08  &--0.12  &0.076  &40\\
    14640$^{a}$ &--0.04  &0.01  &0.05  &--0.05  &0.04  &0.05  &0.04  &0.09  &--0.05  &0.052 &74\\
    14640$^{b}$ &--0.11  &0.05  &0.03  &0.01  &0.08  &0.06  &0.03  &--0.04	  &--0.06  &0.054  &73\\
     15929            &--0.22  &0.07  &--0.03  &--0.11  &0.07  &0.04  &0.08  &0.21  &--0.11  &0.102  &13\\
     21158            &0.11  &--0.01  &0.02  &--0.02  &--0.01  &--0.02  &--0.0	 &0.04  &0.00  &0.022  &95\\
     22002            &0.15  &0.15  &--0.01  &0.05  &0.14  &0.02  &0.01  &0.06  &0.06  &0.073  &44\\
     24232            &0.08  &--0.04  &0.05  &--0.06  &0.02  &0.00  &0.04  &0.05  &--0.05  &0.038  &88\\
     25358            &0.22  &0.00  &--0.02  &--0.04  &0.01	 &0.05  &--0.00  &--0.09  &0.02  &0.053  &73\\
     26744            &--0.00  &0.01  &0.05  &--0.05  &0.04  &0.06  &0.06  &0.10	  &--0.04  &0.050  &78\\
     30344            &--0.02  &--0.11  &--0.08  &--0.11  &--0.02  &--0.03  &--0.01  &0.04  &--0.03  &0.055  &71\\
     33275            &--0.07  &--0.06  &0.02  &--0.09  &0.01  &0.02  &0.02  &0.07  &--0.05  &0.048  &80\\
     33685            &--0.04  &--0.34  &0.02  &--0.11  &0.08  &0.05  &--0.03	  &0.46  &0.08  &0.135  &5\\
     40317$^{a}$ &--0.01  &--0.07  &--0.06  &--0.03  &--0.04  &--0.01	 &0.01  &0.04  &--0.03  &0.036  &89\\
     40317$^{b}$ &--0.05  &--0.07  &--0.08  &--0.07  &0.02  &--0.02  &--0.04  &--0.03  &--0.01  &0.041  &86\\
     48062$^{a}$ &--0.18  &--0.08	 &--0.03  &--0.06  &0.04  &0.03  &0.03  &0.12  &--0.07  &0.068  &52\\
     48062$^{b}$ &--0.23  &0.04  &0.03  &--0.05  &0.10  &0.06  &0.04  &--0.00  &--0.11  &0.078  &37\\
     51581             &--0.00  &0.15  &0.05  &0.06  &0.11  &0.06  &--0.01  &0.04  &0.04  &0.064  &59\\
     53921             &--0.11  &0.03  &0.03  &--0.07  &0.05  &0.08  &--0.01  &0.00  &--0.06  &0.051  &76\\
     56798             &0.03  &0.13  &0.03  &0.08  &0.14  &0.06  &0.00  &0.13  &0.054  &0.068  &52\\
     58968             &--0.03  &--0.08  &--0.04  &--0.13	 &0.06  &0.11  &0.09	 &0.15  &--0.02  &0.076  &39\\
     59291             &--0.06  &0.02  &0.04  &0.01  &0.13  &0.10  &0.07  &0.19  &--0.06  &0.081  &32\\
     60678             &--0.27  &0.02  &0.10  &--0.15  &0.10  &0.05  &0.06 	&0.12  &--0.03   &0.110  &10\\
     73600             &0.04  &--0.06  &--0.06  &--0.04  &0.02  &0.06  &--0.00  &0.04  &--0.04  &0.042  &85\\
     76300             &0.10  &--0.05	 &--0.09  &--0.08	 &--0.02  &--0.03  &0.02  &0.07  &0.01  &0.061  &63\\
     89792             &--0.01  &--0.01  &0.11  &--0.10 	&0.12  &0.00  &0.01  &--0.10  &--0.08  &0.071  &47\\
     89825             &0.22  &0.03  &0.67  &0.29  &0.23  &0.02  &--0.00  &--0.11  &0.08  &0.240  &4\\
     93190             &0.17  &0.10  &0.04  &--0.14  &0.07	 &0.03  &0.05  &0.01	 &--0.00  &0.062  &61\\
   101137             &--0.15  &0.07  &0.19  &--0.10  &0.18  &0.08  &0.15	 &0.07  &--0.09  &0.141  &5\\
   101911             &--0.03  &0.04	 &--0.01  &0.09  &0.12  &--0.01  &0.09	  &0.05  &0.07  &0.056  &70\\
   103738             &--0.09  &0.04  &--0.07  &--0.11	 &0.04  &--0.02  &--0.04  &0.04  &--0.07  &0.049  &78\\
   107528             &0.05  &0.04  &0.03  &--0.08  &0.08	 &0.05  &0.04 &0.04	 &--0.00  &0.042  &85\\
   112584             &--0.12  &--0.18  &--0.08  &--0.16  &--0.08  &--0.01  &--0.01  &0.05	&--0.10  &0.078  &37\\
   115100             &0.28  &0.06  &0.20  &0.06  &0.07  &--0.01  &--0.02  &0.02  &0.05  &0.055  &71\\
           \hline
\end{tabular}
\tablefoot{The first column gives identification of each stars. The second column gives Mg abundances. Abundances of seven element (Na, Mg, Al, Si, Ca, Ti, Ni, Na, Al) relative to Fe are listed in columns 3 to 10. The chemical difference between the sibling candidates and Sun and probability that two stars are born in the same cluster based on chemical difference are given in column 11 and 12, respectively. $^{a}$ and $^{b}$ have the same meanings as in Table~\ref{csp}.}
\end{tiny}
\end{table*}

\subsection{Errors in elemental abundances}

There many possible sources of uncertainty in our derived abundances. These can include continuum placement, line blending and errors in stellar parameters and in atomic data ($\log gf$). Since we performed a differential abundance analysis relative to the Sun, errors due to uncertainties in the $\log gf$ values cancel to first order. Errors due to continuum placement and line blending are estimated by SME. SME gives us a typical error less than 0.01 dex. In section 4.4.2 the uncertainties of stellar parameters were estimated to be $\sigma_{T_{\rm eff}} = 40$ K, $\sigma_{\rm{log} g}=0.06$ dex, and $\sigma_{\rm[Fe/H]}$ = 0.03 dex. The uncertainties in the elemental abundances  associated with these, for three stars (Sun, $\epsilon$ Eri, and HIP 51581), are given in Table~\ref{aberr}. We found that errors in the elemental abundances do not correlate with the errors on the parameters. The total uncertainty was therefore derived by taking the square root of the quadratic sum of the different errors. The average values of the total uncertainties for all elements are between 0.03 and 0.05 dex.

\begin{table*}[htbp]
\centering
\setlength{\tabcolsep}{6pt}
\begin{tiny}
\caption{Errors in the abundances due to the uncertainties in stellar parameters: $T_{\rm eff}$$\pm$40 K, $\log g$$\pm$0.06 dex, [Fe/H]$\pm0.03$ dex. $\Delta_{\rm tot}$ is the total uncertainty.}
\label{aberr}
\begin{tabular}{lrrrrrrrrrrrrrrrrrr}
	\hline
      \hline
 Element    &\multicolumn{3}{c}{$\epsilon$Eri}   &  &&  &\multicolumn{3}{c}{HIP51581}    &  &&    &\multicolumn{3}{c}{Sun}  \\
                    &$\sigma T_{\rm eff}$  &$\sigma$$\log g$  &$\sigma$ [Fe/H] &$\Delta_{\rm tot}$  && &$\sigma T_{\rm eff}$  &$\sigma$$\log g$  &$\sigma$ [Fe/H]  &$\Delta_{\rm tot}$  &&  &$\sigma T_{\rm eff}$  &$\sigma$$\log g$  &$\sigma$ [Fe/H] &$\Delta_{\rm tot}$ \\
      \hline
 $\Delta$[Na/Fe]     &$\pm$0.03   &$\mp$0.01  &$\mp$0.03  &$\pm$0.04  &&  &$\pm$0.02  &0.00  &$\mp$0.03  &$\pm$0.04  &&  &$\pm$0.02  &0.00  &$\mp$0.03  &$\pm$0.04\\
 $\Delta$[Mg/Fe]     &$\pm$0.03  &$\mp$0.03  &$\mp$0.02  &$\pm$0.05  &&  &$\pm$0.03  &$\mp$0.01  &$\mp$0.03  &$\pm$0.04  &&  &$\pm$0.03  &$\mp$0.01  &$\mp$0.02  &$\pm$0.04\\
 $\Delta$[Al/Fe]       &$\pm$0.02  &0.00  &$\mp$0.03  &$\pm$0.04  &&  &$\pm$0.02  &0.00  &$\mp$0.03  &$\pm$0.04  &&  &$\pm$0.02  &0.00  &$\mp$0.03  &$\pm$0.04   \\
 $\Delta$[Si/Fe]       &$\mp$0.01  &$\pm$0.01  &$\mp$0.02  &$\pm$0.02  &&  &$\pm$0.01  &0.00  &$\mp$0.03  &$\pm$0.03  &&  &$\pm$0.01  &0.00  &$\mp$0.03  &$\pm$0.03\\
 $\Delta$[Ca/Fe]     &$\pm$0.05  &$\mp$0.02  &$\mp$0.02  &$\pm$0.06  &&  &$\pm$0.03  &$\mp$0.01  &$\mp$0.04  &$\pm$0.05  &&  &$\pm$0.03  &$\mp$0.01  &$\mp$0.03  &$\pm$0.04\\
 $\Delta$[Ti/Fe]       &$\pm$0.03  &0.00  &$\mp$0.02  &$\pm$0.04  &&  &$\pm$0.03  &0.00  &$\mp$0.04  &$\pm$0.05  &&  &$\pm$0.03  &$\pm$0.01  &$\mp$0.03  &$\pm$0.04 \\
 $\Delta$[Cr/Fe]      &$\pm$0.03  &0.00  &$\mp$0.02  &$\pm$0.04  &&  &$\pm$0.02  &0.00  &$\mp$0.02  &$\pm$0.03  &&  &$\pm$0.01  &$\mp$0.01  &$\mp$0.03  &$\pm$0.03        \\
 $\Delta$[Ni/Fe]      &$\pm$0.01  &$\pm$0.01  &$\mp$0.02  &$\pm$0.02  &&  &$\pm$0.02  &0.00  &$\mp$0.03  &$\pm$0.04  &&  &$\pm$0.02  &0.00  &$\mp$0.03  &$\pm$0.04\\
      \hline
\end{tabular}
\end{tiny}
\end{table*}


\section{Chemical tagging}

Chemical tagging is potentially a powerful tracer of dispersed substructures of the Galactic disk \citep{2002ARA&A..40..487F, 2009PASA...26...11D}. Here we used a chemical tagging method introduced by \cite{2013MNRAS.428.2321M}. This method has been developed to find dispersed clusters in large-scale surveys. Here we give a brief summary of the method. A metric ($\delta_{C}$) was defined as
\begin{equation}
\delta_{C} = \sum_{C}^{N_{C}} \omega_{C} \frac{|A_{C}^{i}-A_{C}^{j}|}{N_{C}},
\label{ct}
\end{equation}
where $N_{C}$ is the number of measured abundances, $A_{C}^{i}$ and $A_{C}^{j}$ are individual abundance ratios of element $C$ with respect to Fe relative to solar for star $i$ and $j$, respectively. $A_{C}$ is the ratio of Fe to H when element $C$ is Fe. As \cite{2013MNRAS.428.2321M} recommended, $\omega_{C}$ that represents a weighting factor for an individual species was fixed at unity. $\delta_{C}$ is the mean absolute difference between any two stars cross all measured elements. The probability that a particular pair of stars are members of the same cluster based on their $\delta_{C}$ can be estimated from a empirical probability function ($P_{\delta_{C}}$, see \cite{2013MNRAS.428.2321M} for more details). \cite{2013MNRAS.428.2321M} suggested a method to verify a group of potential coeval stars from large data sets. We use a similar procedure adapted to our special case. Firstly, $\delta_{C}$ and $P_{\delta_{C}}$ listed in Table~6 were calculated based on 9 elements (Na, Mg, Al, Si, Ca, Ti, Cr, Ni, Fe) between any one solar sibling candidate and the Sun, because the Sun is a standard star in this work and is assumed to have come from a dissolved open cluster. We picked up all calculated pairs with a probability greater than a given confidence limit. The high confidence limit $P_{lim} =$ 85 \% is set in order to reduce contamination stars from other clusters. Secondly, all left sibling candidates making up the pairs each other from above step were re-evaluated. The pairs of two stars which have probability $P_{\delta_{C}}< 85$ \% were cut off in this step. Finally, as \cite{2013MNRAS.428.2321M} suggested, the cluster detection confidence, $P_{clus}$, can be evaluated using the mean of $\delta_{C}$ for the sibling candidates that remain. 

Five potential solar siblings were found in this way, HIP 21158, HIP 24232, HIP 40317, HIP 73600 and HIP 107528, have $\delta_{C} \leq$ 0.042. For this group of stars, pairs of any two stars were made and their $\delta_{C}$ and $P_{\delta_{C}}$ were re-evaluated. Four sibling candidates (HIP 21158, HIP 24232, HIP 40317, and HIP 73600) were identified as cluster stars. This means that 5 stars, including the Sun, might come from a dissolved cluster. This is consistent with the expected mean number of stars ($\sim$ 4.8) in each group that can be detected by given the confidence limit $P_{lim} =$ 85 per cent \citep{2013MNRAS.428.2321M}. Finally, the cluster detection confidence is $P_{clue}=$91 per cent which corresponds to the mean of $\delta_{C}=$0.034.

\section{Potential solar siblings}

Twelve of our stars have iron abundance consistent with the solar value, with systematic and random uncertainties  in [Fe/H]. Comparing with the age of the Sun ($\sim$ 4.6 Gyr) and relevant isochrones, it turns out that 4 out of these 12 stars, HIP 10786, HIP 21158, HIP 40317, and HIP 51581, are consistent with the solar age within 1$\sigma$. They are thus highly potential candidates.

\begin{figure*}[htbp]
\begin{center}
\includegraphics[scale=0.44]{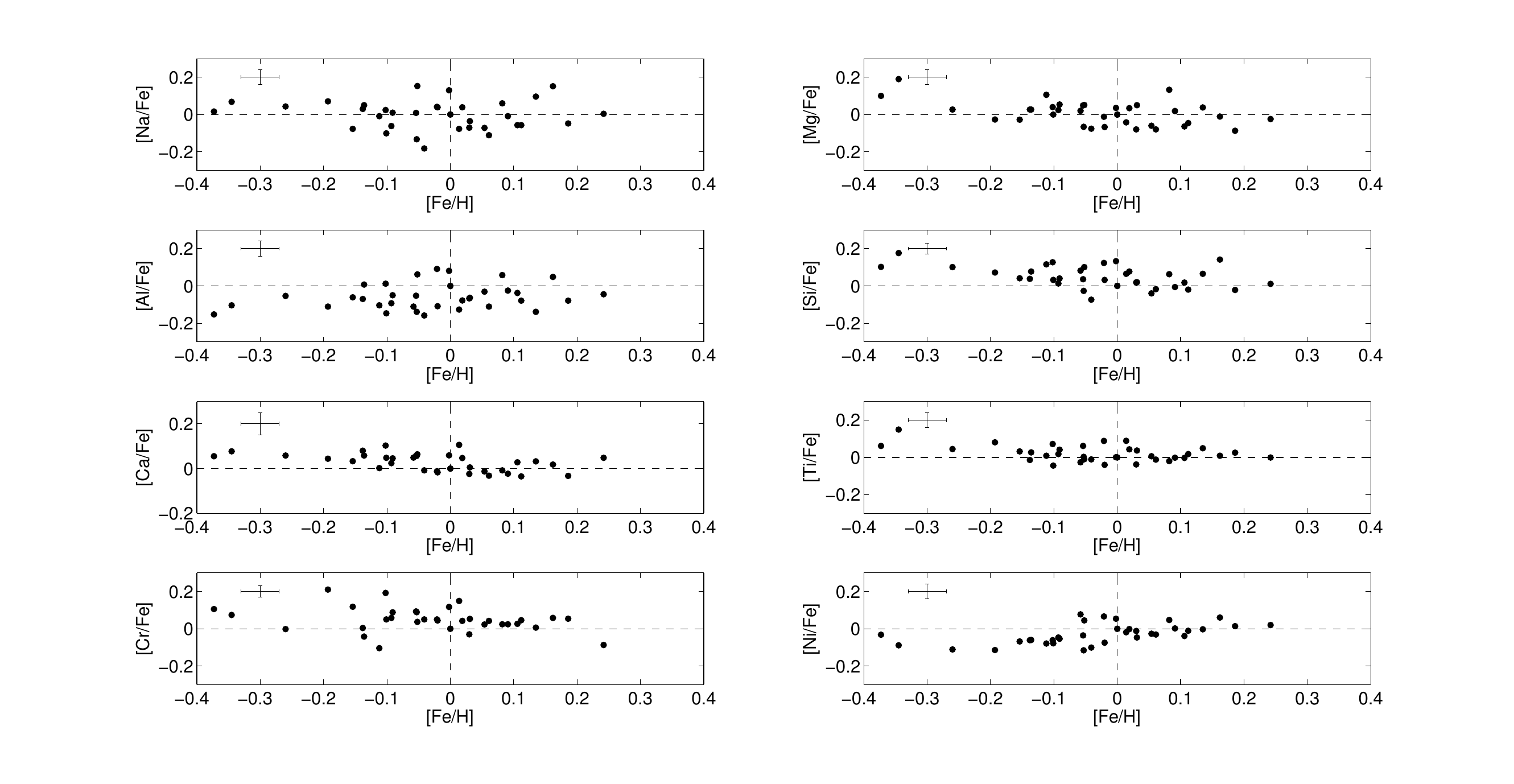}
\caption{Elemental abundance ratios [X/Fe] relative to Fe. Dash lines indicate solar values.}
\label{abunp}
\end{center}
\end{figure*}

In addition to the constraints from [Fe/H] and stellar age, chemical signatures can help us to explore the probability that we have found true solar siblings. The members of a stellar aggregate formed in a common protocluster are found to have a high level of chemical homogeneity \citep{2007AJ....133..694D, 2007AJ....133.1161D, 2010A&A...511A..56P}. The abundance ratios with respect to Fe of all elements, as a function of [Fe/H], are shown in Fig. \ref{abunp}. We found flat trends for the $\alpha$-elements with abundances close to the solar abundances, however, [Al/Fe] and [Ni/Fe] show an increasing abundance for [Fe/H] $<$ 0. We found that 15 out of 32 stars have $P_{\delta_{C}}$ larger than 68 \% according to the chemical tagging method. Only four sibling candidates, HIP 21158, HIP 24232, HIP 40317, and HIP 73600, are tagged by our method with the cluster detection confidence larger than 90\%.

In these two sets of potential siblings, two key targets HIP 21158 and HIP 40317, have ages and abundances in [X/Fe] similar to the Sun. This is consistent with the prediction of \cite{2013MNRAS.428.2321M} that 50 per cent of detections with members from other star aggregates give a high level confidence limit ($P_{lim} =$ 85). It should not be surprising that two of four tagged cluster stars are not solar siblings.

Although our results on [Fe/H] and stellar age support that two stars, HIP 10786 and HIP 51581, are possible solar siblings, they might not have formed from the same protocluster gas as the Sun because of the low probability (< 68 \%) of a member of star aggregate (see Table~\ref{abun}). HIP 24232 and HIP 73600 were tagged as a cluster star in Sect. 5 by the chemical tagging. However, the stellar ages of the stars are at least 2 Gyr younger than that of the Sun. According to the location of these two stars in H--R diagram, HIP 24232 is a typical subgiant star and HIP 73600 is a main sequence star close to the turnoff point. The determined young ages could be trusted. For these two stars we also estimated an age of 1.6$\pm$0.6 Gyr (HIP 24232) and 2.0$\pm$1.6 Gyr (HIP 73600) by using the PARAM database \citep{2006A&A...458..609D}. It implies that they might be the members of a later dissolved cluster which has solar abundances. 

One key target (HIP 21158) has been mentioned by both \cite{2010MNRAS.407..458B} and \cite{2012NewA...17..514B} as a potential solar sibling. Although HIP 21158 is tagged as a cluster star based on [X/Fe] abundance ratios, it appears to have +0.1 offset in [Mg/H] and metallicity. \cite{2014ApJ...787..154R} also found that it has super-solar abundances for all their measured elements. The inconsistency in HIP 21158 might imply that [Fe/H] should be given a larger weighting factor in Eq.~\ref{ct} rather than 1. The age derived for HIP 21158 (5.2 Gyr) is slightly older than the Sun and consistent with several other studies on the stellar ages \citep{2001A&A...377..911F, 2007ApJS..168..297T, 2011A&A...530A.138C}. Another key target, HIP 40317, has perfectly solar abundances both in [X/Fe] and [X/H] ratios. Two stellar ages were obtained for this star, because two observations were made using two different telescopes. Although the stellar ages suffer large uncertainty because it is a main sequence star, the mean difference in age between this star and the Sun is less than 1.2 Gyr. This suggests that it could be a lost sibling of the Sun.

\section{Properties and dynamics of the solar siblings}

\subsection {Rotational velocity of star}
According to the previous studies of stellar rotational revolution, the decline of rotation with age caused by angular momentum loss through the ionized wind is well established from observations of clusters \citep{2010ARA&A..48..581S, 2008ApJ...687.1264M}. Nearly all of the velocities of F, G and K stars fall below 12 km $\rm s^{-1}$ \citep{1987ApJ...321..459R} by the age of the Hyades ($\sim$625 Myr; \cite{1998A&A...331...81P}). If we assume that all solar sibling candidates are born in the same parent cluster, they might have more or less the same rotation rate as that of the Sun after about 4.6 Gyr declining of rotation. Since the projected rotational velocity of HIP 40317 from our synthesis fitting is 0.8 $\mathrm{km~s}^{-1}$, the star could very well have the same rotational velocity as the Sun (depending on inclination $\sin i$).

\subsection{Radial Velocity of the Sun's siblings}

\begin{table}
 \centering
  \caption{Initial conditions of the Sun's birth cluster}
  \label{ic}
  \begin{tabular}{@{} c c c c}  \hline \hline
   $M{\rm _c}$  & $R{\rm _c}$  & $N$ &$\sigma \rm_v $  \\ 
   ($M_{\odot}$) &(pc) & &($\mathrm{km~s}^{-1}$) \\
   \hline 
   507.5 & 2 & 875 &0.8 \\
  510.3 & 0.5 & 875&1.6\\
  525.8  &1.5 & 875 & 0.9\\
  549.8 &1 & 875 & 1.1 \\
 804.6 & 3 & 1500 & 0.8\\
 1023.6 & 2 & 1741& 1.1\\
 1056.6 &2.5 & 1741& 1.0\\
 1067.7 &1.5  & 1740& 1.3\\
 1125.0 & 2 & 1742& 1.2\\ \hline	
\end{tabular}
\end{table} 

The solar abundances and age of HIP 40317 make this star a highly potential candidate to be one of the lost siblings of the Sun, however, the barycentric radial velocity of HIP40317 is equal to 34.2 $\mathrm{km~s^{-1}}$.  By performing N-body simulations we study the evolution in the Galaxy of the already extinct Sun's birth cluster. Our aim is to conclude whether or not nearby solar siblings may exhibit high radial velocities. At the beginning of the simulations the parental cluster of the Sun obeys to a spherical Plummer density distribution function \citep{1911MNRAS..71..460P} together with a Kroupa IMF \citep{2001MNRAS.322..231K}. The initial mass $(M_\mathrm{c})$ and radius $(R_\mathrm{c})$ of the Sun's birth cluster were set according to the values suggested by \cite{2009ApJ...696L..13P}. The initial conditions used in the simulations are listed in Table~\ref{ic}.  

\begin{table}
 \centering
 \begin{minipage}{80mm}
  \caption{Galactic parameters of the Milky Way}
  \label{gp}
  \begin{tabular}{@{} c c}  \hline
 \multicolumn{2}{c}{\textbf{ Axisymmetric component}} \\ \hline 
Mass of the bulge($M_\mathrm{b}$) & $1.41\times 10^{10}$ $\mathrm{M}_{\odot}$  \\ 
Scale length bulge ($b_1$) & 0.3873 kpc\\
Disk mass ($M_\mathrm{d}$) &  $8.56\times10^{10}$ $\mathrm{M}_{\odot}$\\
Scale length disk 1 ($a_2$) & 5.31 kpc\\
Scale length disk 2 ($b_2$) & 0.25\\
Halo mass ($M_\mathrm{h}$) & $1.07\times 10^{11} $ $\mathrm{M}_{\odot}$\\
Scale length halo ($a_3$) & 12 kpc\\  \hline 
 \multicolumn{2}{c}{\textbf{Central Bar} } \\ \hline 
Pattern speed ($\Omega_\mathrm{bar}$) & 40--70 $\mathrm{km~s}^{-1}\mathrm{kpc}^{-1}$\\ 
Semi-major axis ($a$) & 3.12 kpc \\
Axis ratio ($b/a$) & 0.37 \\
Mass ($M_\mathrm{{bar}}$) & $9.8\times10^9$--$1.1\times10^{10}$ $\mathrm{M}_{\odot}$\\  \hline
 \multicolumn{2}{c}{\textbf{ Spiral arms}} \\ \hline 
Pattern speed ($\Omega_\mathrm{sp}$) & 15--30  $\mathrm{km~s}^{-1}\mathrm{kpc}^{-1}$\\
Locus beginning ($R_\mathrm{sp}$) & 3.12 kpc\\
Number of spiral arms ($m$) & 2, 4\\
Spiral amplitude ($A_\mathrm{sp}$) &  650--1100 $[\mathrm{km~s}^{-1}]^2\mathrm{kpc}^{-1}$\\
Pitch angle ($i$) & 12.8$^o$\\
Scale length ($R_\mathrm{\Sigma}$) & 2.5 kpc\\  \hline
\end{tabular}
\end{minipage}
\end{table} 

The Milky Way was modeled as an analytical potential consisting of an axisymmetric component together with a central bar and spiral arms. We adopted the same Galactic model as \cite{2014arXiv1410.2238M} where the axisymmetric part of the Galaxy is modelled by using the potential of \cite{1991RMxAA..22..255A} which consist of a bulge, disk and a dark matter halo. The parameters that describe the axisymmetric component of the Milky Way are listed in Table~\ref{gp}.

The central bar of the Milky Way was modeled with a Ferrers potential \citep{1877pam..14..1}, which is described by a density distribution of the form:
\begin{equation}
\mathrm{\rho_\mathrm{bar}= } 
\begin{cases}
\rho_0 \left( 1-n^2 \right)^k, & n \leq 1\\
0 & n \geq 1,
\end{cases}  
\end{equation}
where $\rho_0$ represents the central density of the bar which is related to its mass $M_\mathrm{bar}$. $n^2$ determines the shape of the potential of the bar. On the Galactic plane, $n^2= x^2/a^2 + y^2/b^2$, being $a$ and $b$ its semi-major and semi-minor axis respectively. The parameter $k$ measures the degree of concentration of the bar. In the simulations, we use $k=1$ \citep{2011MNRAS.418.1176R}. In addition to the former parameters,  we assume that the bar rotates as a rigid body with constant pattern speed $\Omega_\mathrm{bar}$. The values of the mass, semi-major axis, axis ratio and pattern speed of the bar are listed in Table~\ref{gp} and they were set to fit the observations made by the \textit{COBE/DIRBE} survey \citep[see][]{2004ApJ...609..144P, 2011MNRAS.418.1176R, 2012AJ....143...73P}.

The spiral structure of the Milky Way is represented as a periodic perturbation of the axisymmetric component of the Galaxy. The potential associated to this perturbation is given by:
\begin{equation}
\phi_{sp}= -A_{sp}Re^{-R/R_{\Sigma}}\cos{\left(m(\phi)-g(R) \right) }, \label{speq} \,
\end{equation} 
where $A_\mathrm{sp}$ is the amplitude of the spiral arms,  $R$ and $\phi$ are the galactocentric cylindrical coordinates,  $R_\mathrm{\Sigma}$ and $m$ are the scale length and number of  spiral arms respectively and $g(R)$ is the function that defines the locus shape of spiral arms. We use the same shape factor as \cite{2011MNRAS.418.1423A}:
\begin{equation}
g(R)= \left(  \frac{m}{\xi \tan{i}} \right) \ln{\left( 1+ \left( \frac{R}{R_\mathrm{sp}} \right)^\xi  \right)},
\end{equation}
where $\xi$ is a parameter that measures how sharply occurs the change from a bar to a spiral structure in the inner regions.  $\xi \rightarrow \infty$ produces spiral arms that begin forming an angle of $\sim 90^o$ with the line that joins the two starting points of the locus, thus we chose $\xi=100$ \citep{2011MNRAS.418.1423A}. $R_\mathrm{sp}$ is the separation distance of the beginning of the spiral shape locus and $\tan{i}$ is the tangent of the pitch angle. Additionally, we assume that the spiral arms of the Galaxy rotate as a rigid body with pattern speed $\mathrm{\Omega_{sp}}$. The values of the former parameters are listed in Table~\ref{gp} and they correspond to the best fit  to the Perseus and Scutum arms of the Milky Way \citep[see][]{2011MNRAS.418.1423A}. 

In the numerical simulations the Sun's birth cluster is evolved under the influence of its self-gravity, stellar evolution and the external gravitational field generated by the analytical model of the Galaxy. The motion of the stars due to their self-gravity was computed by the \textit{HUAYNO} code \citep{2012NewA...17..711P}. The motion of the stars under the external tidal field of the Milky Way was computed by a 6th-order rotating \textit{BRIDGE} \citep{2014arXiv1410.2238M}. Additionally, we used the \textit{SeBa} code\citep{1996A&A...309..179P, 2012A&A...546A..70T} to model the stellar evolution of the stars. We assumed a solar metallicity ($Z= 0.02$ or $\rm{[Fe/H]}= 0$) for the Sun's birth cluster. \textit{HUAYNO}, the 6th-order rotating \textit{BRIDGE} and \textit{SeBa} were coupled through the \textit{AMUSE} framework \citep{2013CoPhC.183..456P}.

The initial phase-space coordinates of the Sun's birth cluster center of mass ($x_\mathrm{cm}$, $y_\mathrm{cm}$, $v_\mathrm{x_{cm}}$, $v_\mathrm{y_{cm}}$) were obtained by evolving the orbit of the Sun backwards in time taking into account the uncertainty in the Sun's current position and velocity, as is shown in \cite{2014arXiv1410.2238M}. The orbit integration backwards in time gives a distribution of all the possible positions and velocities of the Sun at its birth, we therefore choose one position and velocity from this distribution to be the initial phase-space coordinates of the Sun's birth cluster center of mass. This procedure was done for different bar and spiral arms parameters.  

\begin{figure}
\centering
\includegraphics[width= 8cm, height= 14cm]{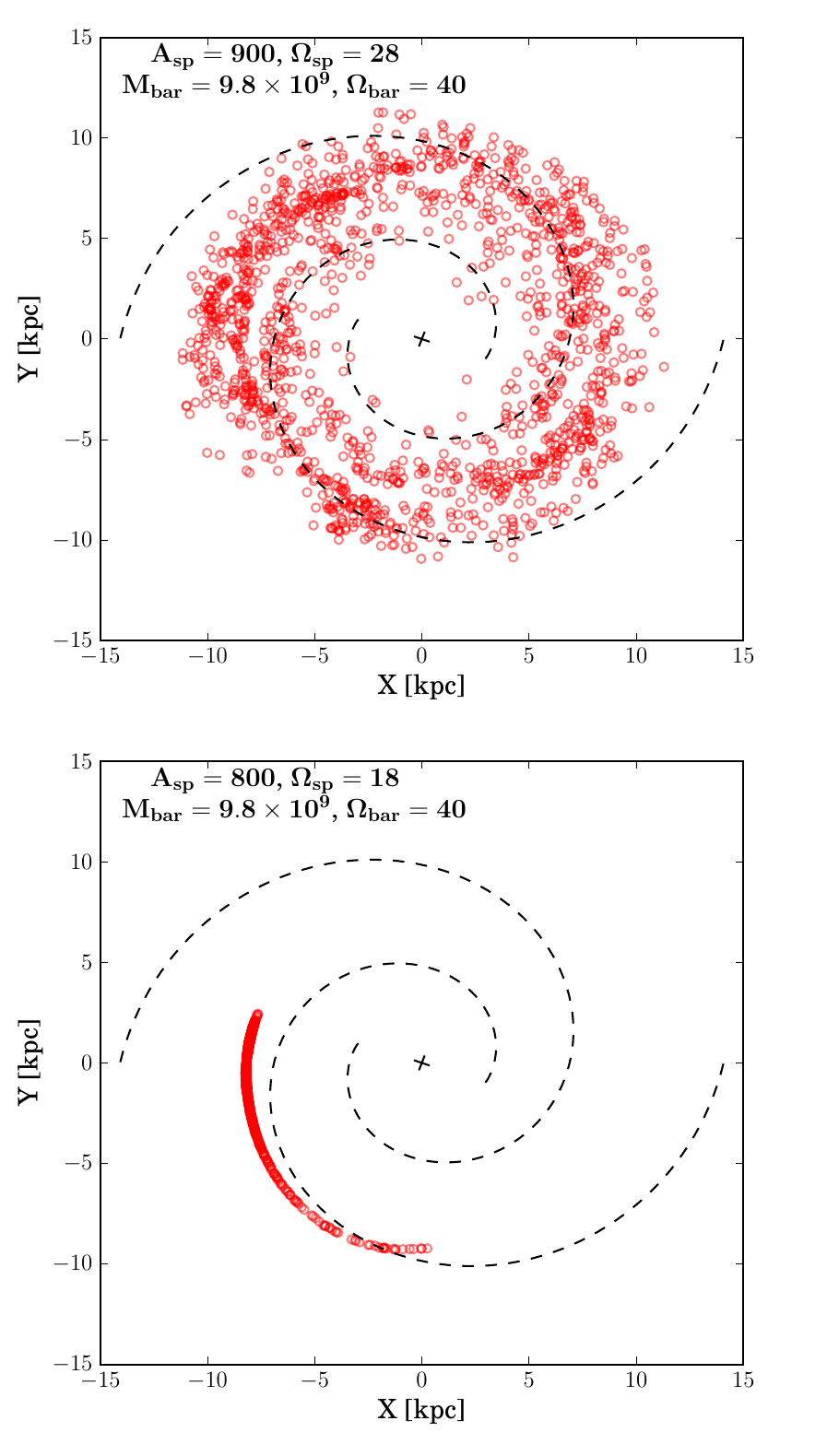}\\
\caption{Final distribution on the Galactic plane of solar siblings when $M_\mathrm{c}= 804.6$ $M_\odot$ and $R_\mathrm{c}= 3$ $\mathrm{pc}$. Note that the final phase-space coordinates of the solar siblings depend on the configuration of the Galactic potential. \textbf{Top:} The Sun's siblings are dispersed on the Galactic disk. \textbf{Bottom:} The solar siblings are located in a specific region on the Galactic disk. The dashed black lines represent the potential of the spiral arms at the end of the simulation.}
\label{dis}
\end{figure}

Once the Sun's birth cluster is located at coordinates ($x_\mathrm{cm}$, $y_\mathrm{cm}$, $v_\mathrm{x_{cm}}$, $v_\mathrm{y_{cm}}$), it is evolved forwards in time during 4.6 Gyr. We used a time step of 0.5 Myr and 0.16 Myr for \textit{HUAYNO} and the 6th-order rotating \textit{BRIDGE}, respectively. These values give an maximum energy error of $10^{-7}$ during the entire simulation.  We carried out 1071 simulations in total assuming different bar and spiral arm parameters in the Galactic model as well as different initial masses and radii of the Sun's birth cluster.

Depending on a given combination of bar and spiral arms parameters, the current distribution on the $xy$ plane of the solar siblings could be highly dispersed or not, as can be seen in Fig.~\ref{dis}. Since all the bar and spiral arms parameters listed in Table \ref{gp} are equally probable, the final distribution of the solar siblings shown in this figure is equally plausible. In the case of a highly dispersed distribution, the difference between the maximum and minimum position of the solar siblings ($\Delta R$) can be higher than $3$~kpc. Additionally, the siblings of the Sun may exhibit a broad range of azimuths. An example of a high dispersed distribution of solar siblings is shown at the top panel of Fig.~\ref{dis}. The set of Galactic parameters that produce high dispersion on the current distribution of the solar siblings are:
\begin{itemize}
\item When $m=2$: $27\leq \mathrm{\Omega_{sp}} \leq 28$ $\mathrm{km~s}^{-1}\mathrm{kpc}^{-1}$ and $A_\mathrm{sp}\geq 900 $ $[\mathrm{km~s}^{-1}]^2\mathrm{kpc}^{-1}$.

\item When $m=4$:  $\Omega_\mathrm{sp} \geq 18$ $\mathrm{km~s}^{-1}\mathrm{kpc}^{-1}$; $\forall$ $A_\mathrm{sp}$. 
\end{itemize}

We found that the high dispersion in the current phase-space coordinates of the solar siblings does not depend on $M_\mathrm{c}$ and $R_\mathrm{c}$.  For the specific case shown at the top panel of Fig.~\ref{dis}, the solar siblings span a range of radii between $2.9$ and  $11.6$ kpc ($\Delta R= 8.6$~kpc). Hereafter we will call the high dispersed distribution of solar siblings as the high dispersion case. 

The current distribution on the $xy$ plane of the solar siblings could also exhibit a small radial and angular dispersion, as can be observed at the bottom panel of Fig.~\ref{dis}. The set of Galactic parameters that produce low dispersion on the current distribution of solar siblings are:
\begin{itemize}
\item All the variations in $M_\mathrm{bar}$ and $\Omega_\mathrm{bar}$ when $A_\mathrm{sp} = 650 ~ [\mathrm{km~s}^{-1}]^2\mathrm{kpc}^{-1}$\, , $\Omega_\mathrm{sp}$= $20$ $\mathrm{km~s}^{-1}\mathrm{kpc}^{-1}$ and $m=2$. 
\item  When $m=2$:  $\mathrm{\Omega_{sp}} \neq 27, 28$ $\mathrm{km~s}^{-1}\mathrm{kpc}^{-1}$. 
\item When $m=4$:  $\mathrm{\Omega_{sp}}= 16$ $\mathrm{km~s}^{-1}\mathrm{kpc}^{-1}$; $\forall$ $A_\mathrm{sp}$.
\end{itemize} 

Hereafter we will call the low dispersed distribution of solar siblings as the low dispersion case.  For the specific set of Galactic parameters shown at the bottom panel of Fig.~\ref{dis}, we found that the radii of the solar siblings  are in the range $8.0 \leq \mathrm{R}\leq 9.3$ kpc ($\Delta R= 1.3$~kpc).

\begin{figure}
\centering
\includegraphics[width= 7.5cm, height= 14cm]{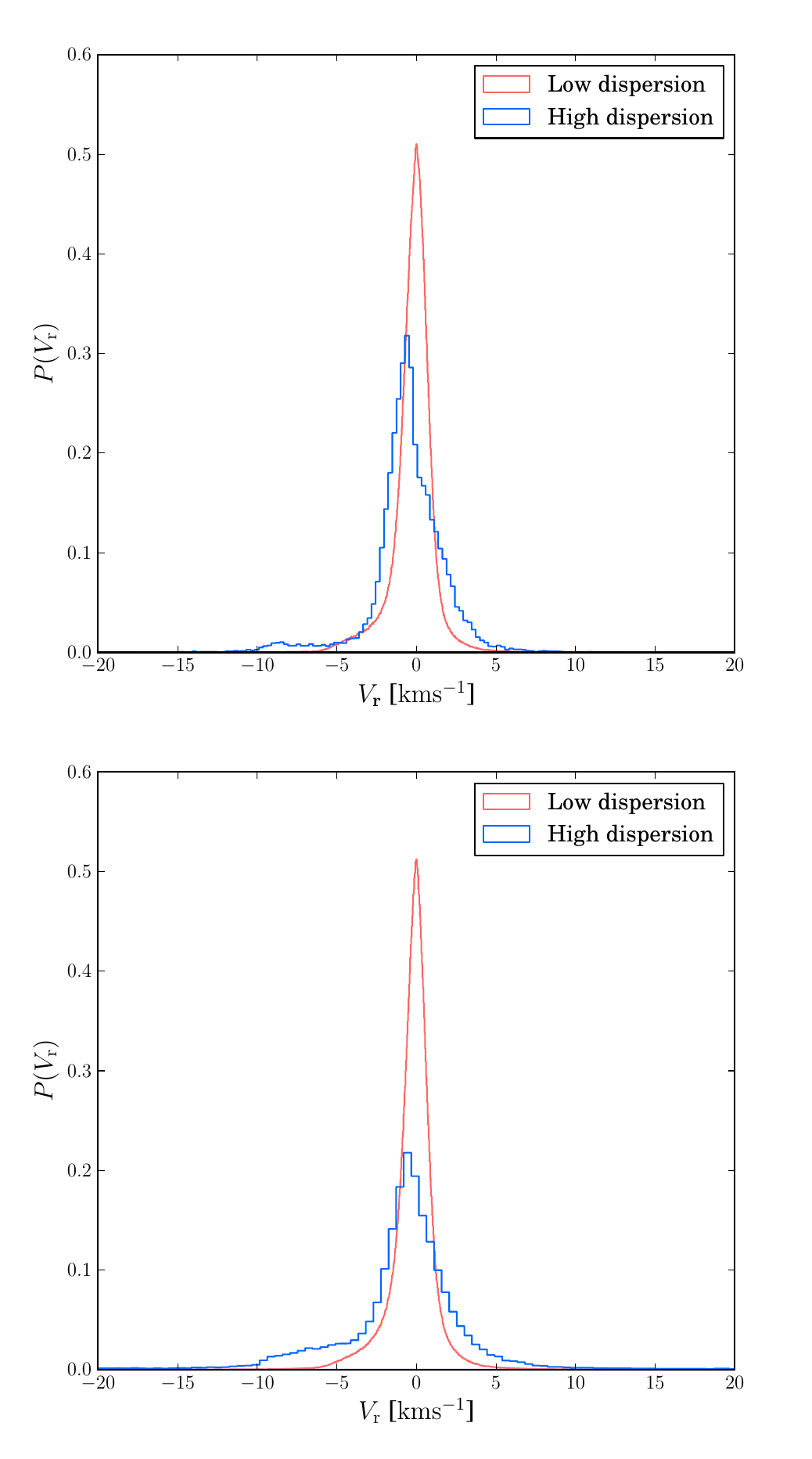}\\
\caption{Distribution of radial velocities $\mathrm{P(V_r)}$ of the solar siblings for the high and low dispersion cases when: \textbf{Top:} The selection criteria of Eq. \eqref{sc} is applied. \textbf{Bottom:} Only the parallax in Eq. \eqref{sc} is taken into account. The initial conditions of the Sun's birth cluster here are:  $\mathrm{M_c}= 1125$ $M_\odot$ and $\mathrm{R_c= 2}$ $\mathrm{pc}$ respectively. }
\label{rvp}
\end{figure}

We computed the astrometric properties of the Sun's siblings such as parallaxes ($\varpi$), proper motions ($\mu$) and radial velocities ($\mathrm{V_r}$) for the cases of high and low dispersion. Given that for one simulation the final distribution of solar siblings could be located all over the Galactic disk (e.g. top panel Fig.~\ref{dis}), we first need to select the stars that have the same galactocentric position as the Sun ($\mathrm{R}= 8.5 \pm 0.5$ kpc). The astrometric properties of the Sun's siblings are then measured with respect to each of those Sun-like stars. We are interested in looking at the radial velocity of nearby solar siblings on almost the same orbit of the Sun. Therefore, \textit{following \cite{2010MNRAS.407..458B} we choose the radial velocity of solar siblings that satisfy selection criteria given by Eq. \eqref{sc}}. This equation makes use of the observationally established value of $(V_\mathrm{LSR}+V_\odot)/R_\odot$ in order to avoid introducing biases related to inadequacies in the simulated phase-space distribution of solar siblings \citep{2010MNRAS.407..458B}. 

However, since the proper motion of the recently discovered solar sibling --HD 162826-- does not correspond to the former selection criteria \citep[see][]{2014ApJ...787..154R}, we also analyze the radial velocities of solar siblings without taking into account their proper motion.  The computation of the astrometric properties of the solar siblings was done by using the Python's package \textit{PyGaia}\footnote{https://pypi.python.org/pypi/PyGaia/0.5}, which is a toolkit for basic Gaia data simulation, manipulation and analysis. 

\begin{figure}
\centering
\includegraphics[width= 8cm, height= 14cm]{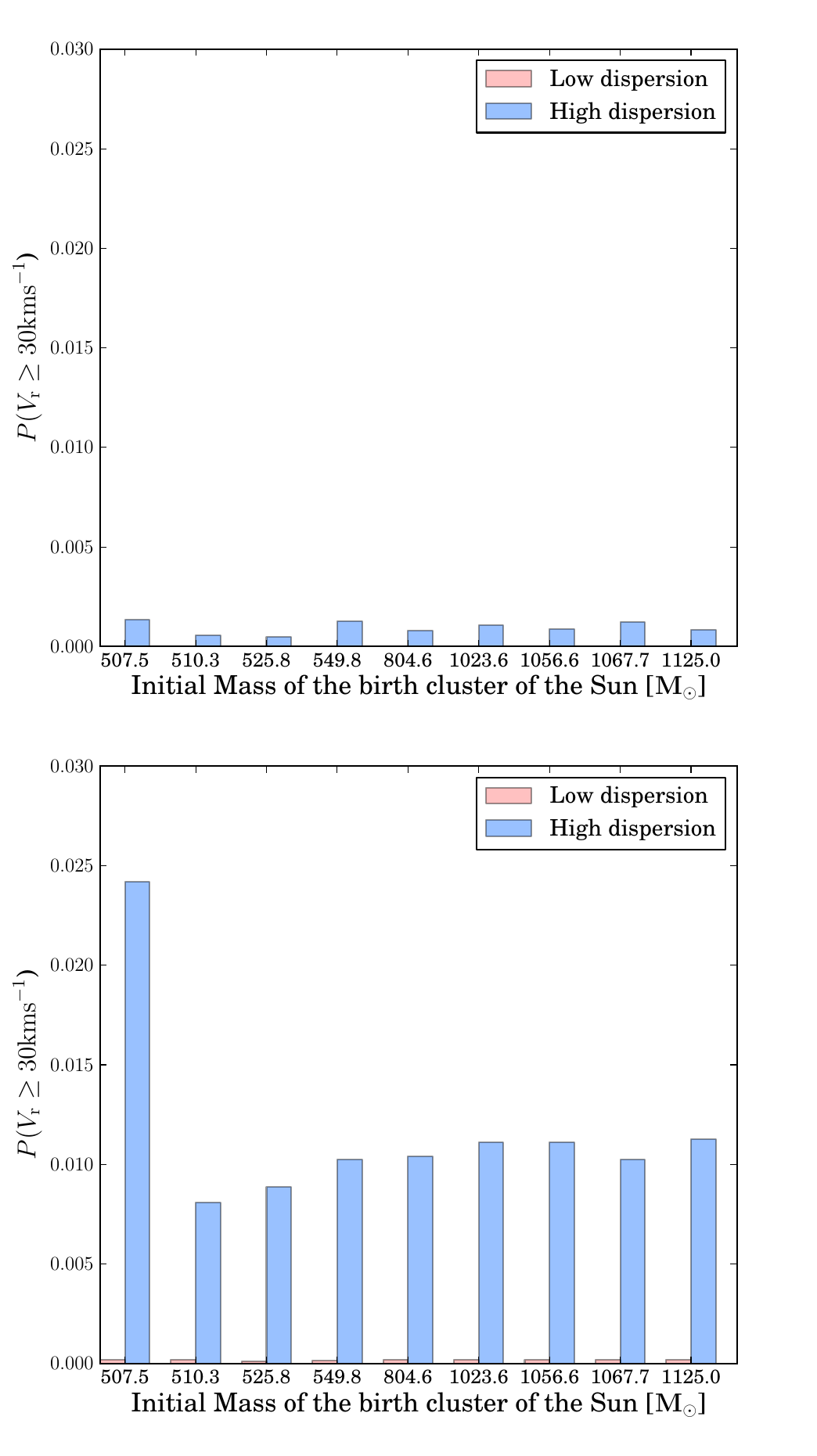}\\
\caption{Probability of finding solar siblings with radial velocity higher than $30~\mathrm{km~s^{-1}}$ in absolute value as a function of the initial mass of the Sun's birth cluster. \textbf{Top:} The selection criteria of Eq. \eqref{sc} is applied. \textbf{Bottom:} Only the parallax in Eq. \eqref{sc} is taken into account.}
\label{his}
\end{figure}

In Fig.~\ref{rvp} we show the distribution of radial velocities $P(V_\mathrm{r})$ of the solar siblings when $M_\mathrm{c}= 1125$ $M_\odot$ and $R_\mathrm{c}= 2$ $\mathrm{pc}$. The velocity distribution were built by considering the Galactic parameters that produce either a high or low dispersion on the final distribution of the Sun's siblings. As can be seen, $P(V_\mathrm{r})$ is peaked at $V_\mathrm{r}\sim 0$ $\mathrm{km~s^{-1}}$ and most of the radial velocities lie in the range $-10 \leq V_\mathrm{r}\leq 10$ $\mathrm{km~s^{-1}}$, regardless of the selection criteria or the Galactic parameters. We found that there is a probability between 97\% and 99\% that the radial velocity of the Sun's siblings lie in the previous range. The distribution shown in Fig.~\ref{rvp} is the same for the initial conditions of the Sun's birth cluster listed in Table~\ref{ic}.

Since the star HIP 40317 has a radial velocity of $34$~km~s$^{-1}$, we computed the probability of finding solar siblings with radial velocities higher than $30$~ $\mathrm{km~s^{-1}}$. The results are shown in Fig.~\ref{his}. Note that when the selection criteria of Eq.~\eqref{sc} is applied (see top panel of Fig.~\ref{his}), such probability is much smaller than 0.5\%. However, if only the parallax in Eq.~\eqref{sc} is taken into account (see bottom panel of Fig.~\ref{his}), the probability of finding solar siblings with high radial velocity could be up to $\sim$ 2.5\% in the case of current distribution of solar siblings highly dispersed on the Galactic disk (see blue bars). According to these results, it is unlikely to find solar siblings with $V_\mathrm{r}\sim 30$~$\mathrm{km~s^{-1}}$.

Comparing with solar neighborhood observations, the estimated range of radial velocity is substantially less than the Galactic velocity dispersion ($\sim$~45~$\mathrm{km~s^{-1}}$) for stars of the solar age \citep[see][]{2009A&A...501..941H}. However, increased velocity dispersion could be explained as a natural consequence of the radial migration of solar age stars \citep{2014RvMP...86....1S} which come from many different birth places rather than that of the Sun.

On the other hand, it has also been shown that the giant molecular clouds (GMCs) could heat the disk stars and the orbits of stellar clusters (Gustafsson et al. 2014, submitted) and they are missed in our models. Past studies have indicated that GMCs on their own are not able to heat the disk to the observed dispersion \citep{2002MNRAS.337..731H}, however, a single star from star cluster might be greatly influenced. If we relax the 3 pc for the virial radius of the proto-cluster, the velocity dispersion in the cluster would be much larger than that of the chosen cluster half-mass radii and the speed of the currently observable solar sibling can be explained easily. However, if we assumed that HD 162826 ($V_\mathrm{r}\sim 2$~$\mathrm{km~s^{-1}}$) discovered by \cite{2014ApJ...787..154R} is a solar sibling, the primordial cluster should have had a smaller viral radius.

Thus, although we found that the abundance and age data are in favor of sibling status for HIP 40317 (see Sect. 6), it is not directly supported by the dynamical arguments. We note that further studies of the dynamics of stars and stellar clusters in increasingly realistic conditions will continue to impact the studies of solar siblings.


\section{Conclusions}

We have obtained high-resolution spectra of 33 out of 57 solar sibling candidates which were selected based on thier colours and constraints in the proper motion and parallax space. Stellar parameters ($T_{\rm eff}$, $\log g$, [Fe/H], $v{\sin}i$) were determined through both a purely spectroscopic approach and partly physical method, respectively. Elemental abundances were determined by comparing observed spectra with synthetic spectra based on the stellar parameters obtained from our partly physical method (see Sec. 4.2.2). In order to calculate errors in elemental abundances, uncertainties of $T_{\rm eff}$, $\log g$ and [Fe/H] were estimated to be about 40 K, 0.06 dex and 0.03 dex based on 5 benchmark stars. Stellar ages were calculated from isochrones by maximizing the probability distribution functions.

Given the constraints on metallicity and stellar age, we found that four stars (HIP 10786, HIP 21158, HIP 40317 and HIP 51581) stand out from our candidate list. They have both metallicity and age close to the solar values within error bars. From an analysis of the Na, Mg, Al, Si, Ca, Ti, Cr and Ni abundances of our observed candidates, we performed chemical tagging to identify cluster stars from the dissolved parent cluster. This resulted in a high probability that four sibling candidates (HIP 21158, HIP 24232, HIP 40317, and HIP 73600) share the same origin as the Sun. However, only HIP 40317 was identified as a possible solar sibling. We also noted that the rotational velocity of HIP 40317 could have the same rotational velocity as the Sun depending on the ${\sin}i$. We performed simulations of the Sun's birth cluster in an analytical model of the Galaxy and found that most of the radial velocities of the solar siblings lie in the range $-10 \leq \mathrm{V_r}\leq 10$ $\mathrm{km~s^{-1}}$, which is smaller than the radial velocity of HIP 40317. We found that a fraction of stars from the star cluster might be accelerated to high velocity by heating sources, however, the probability of high radial velocity solar siblings based on our dynamical analysis is too low to support that star HIP 40317 is a lost sibling of the Sun.

If we assume that HIP 40317 is a solar sibling, it means that only a very small fraction of sibling candidates ($\lesssim$3\%) are actually solar siblings. This is consistent with the prediction that within 100 pc from the Sun, about 1--6 are expected in our sample according to the simulations done by \cite{2009ApJ...696L..13P}. More recently \cite{2014ApJ...787..154R} discovered only one solar sibling amongst 30 candidates, which is very similar to our results. 

This leads to the question, how can we find solar siblings more efficiently. It is not clear what accuracy is needed to distinguish field stars and cluster stars. Since the probabilities of stars which are members of dissolved cluster are estimated based on an empirical function, a chemical tagging experiment of large scale should calibrate against a number of known clusters \citep{2013MNRAS.428.2321M}. More chemical dimensions should be used to probe the formation sites of stars instead of 9 elements. Further, a simple Galactic potential was used to simulate the process of cluster disruption in both \cite{2009ApJ...696L..13P} and \cite{2010MNRAS.407..458B}. It has been argued that solar siblings are unlikely to be found within the solar vicinity because of the influence of the perturbed Galactic gravitational field associated with spiral density waves \citep{2011MNRAS.412.1771M}. More detailed modeling of stellar orbits in a realistic potential could potentially prove more efficient at finding solar siblings.
   
   
\begin{acknowledgements}
       The authors wish to thank the referee, Ivan Ram\'{\i}rez, for thoughtful comments that improved the manuscript. We would like to thank Lennart Lindegren and Ross Church for valuable comments that improved the analysis and text of the paper. This project was completed under GREAT -- ITN network which is funded through the European Union Seventh Framework Programme [FP7/2007-2013] under grant agreement n$^{\rm o}$ 264895. S.F. and G.R. are funded by grant No. 621-2011-5042 from The Swedish Research Council. T.B. is founded by grant No. 621-2009-3911 from The Swedish Research Council. This work also was supported by the Netherlands Research Council NWO (grants \#643.200.503, \#639.073.803 and \#614.061.608) and by the Netherlands Research School for Astronomy (NOVA). This research has made use of the SIMBAD database, operated at CDS, Strasbourg, France.
\end{acknowledgements}

\bibliographystyle{aa.bst} 
\bibliography{cheng-referen} 

\begin{thebibliography}{86}
\expandafter\ifx\csname natexlab\endcsname\relax\def\natexlab#1{#1}\fi

\bibitem[{{Adams}(2010)}]{2010ARA&A..48...47A}
{Adams}, F.~C. 2010, \araa, 48, 47

\bibitem[{{Allen} \& {Santillan}(1991)}]{1991RMxAA..22..255A}
{Allen}, C. \& {Santillan}, A. 1991, \rmxaa, 22, 255

\bibitem[{{Alonso} {et~al.}(1995){Alonso}, {Arribas}, \&
  {Martinez-Roger}}]{1995A&A...297..197A}
{Alonso}, A., {Arribas}, S., \& {Martinez-Roger}, C. 1995, \aap, 297, 197

\bibitem[{{Alonso} {et~al.}(1996{\natexlab{a}}){Alonso}, {Arribas}, \&
  {Martinez-Roger}}]{1996A&AS..117..227A}
{Alonso}, A., {Arribas}, S., \& {Martinez-Roger}, C. 1996{\natexlab{a}}, \aaps,
  117, 227

\bibitem[{{Alonso} {et~al.}(1996{\natexlab{b}}){Alonso}, {Arribas}, \&
  {Martinez-Roger}}]{1996A&A...313..873A}
{Alonso}, A., {Arribas}, S., \& {Martinez-Roger}, C. 1996{\natexlab{b}}, \aap,
  313, 873

\bibitem[{{Antoja} {et~al.}(2011){Antoja}, {Figueras}, {Romero-G{\'o}mez},
  {Pichardo}, {Valenzuela}, \& {Moreno}}]{2011MNRAS.418.1423A}
{Antoja}, T., {Figueras}, F., {Romero-G{\'o}mez}, M., {et~al.} 2011, \mnras,
  418, 1423

\bibitem[{{Aufdenberg} {et~al.}(2005){Aufdenberg}, {Ludwig}, \&
  {Kervella}}]{2005ApJ...633..424A}
{Aufdenberg}, J.~P., {Ludwig}, H.-G., \& {Kervella}, P. 2005, \apj, 633, 424

\bibitem[{{Bagnulo} {et~al.}(2003){Bagnulo}, {Jehin}, {Ledoux}, {Cabanac},
  {Melo}, {Gilmozzi}, \& {ESO Paranal Science Operations
  Team}}]{2003Msngr.114...10B}
{Bagnulo}, S., {Jehin}, E., {Ledoux}, C., {et~al.} 2003, The Messenger, 114, 10

\bibitem[{{Batista} {et~al.}(2014){Batista}, {Adibekyan}, {Sousa}, {Santos},
  {Delgado Mena}, \& {Hakobyan}}]{2014A&A...564A..43B}
{Batista}, S.~F.~A., {Adibekyan}, V.~Z., {Sousa}, S.~G., {et~al.} 2014, \aap,
  564, A43

\bibitem[{{Batista} \& {Fernandes}(2012)}]{2012NewA...17..514B}
{Batista}, S.~F.~A. \& {Fernandes}, J. 2012, \na, 17, 514

\bibitem[{{Bedell} {et~al.}(2014){Bedell}, {Mel{\'e}ndez}, {Bean},
  {Ram{\'{\i}}rez}, {Leite}, \& {Asplund}}]{2014ApJ...795...23B}
{Bedell}, M., {Mel{\'e}ndez}, J., {Bean}, J.~L., {et~al.} 2014, \apj, 795, 23

\bibitem[{{Bensby} {et~al.}(2011){Bensby}, {Ad{\'e}n}, {Mel{\'e}ndez}, {Gould},
  {Feltzing}, {Asplund}, {Johnson}, {Lucatello}, {Yee}, {Ram{\'{\i}}rez},
  {Cohen}, {Thompson}, {Bond}, {Gal-Yam}, {Han}, {Sumi}, {Suzuki}, {Wada},
  {Miyake}, {Furusawa}, {Ohmori}, {Saito}, {Tristram}, \&
  {Bennett}}]{2011A&A...533A.134B}
{Bensby}, T., {Ad{\'e}n}, D., {Mel{\'e}ndez}, J., {et~al.} 2011, \aap, 533,
  A134

\bibitem[{{Bensby} {et~al.}(2003){Bensby}, {Feltzing}, \&
  {Lundstr{\"o}m}}]{2003A&A...410..527B}
{Bensby}, T., {Feltzing}, S., \& {Lundstr{\"o}m}, I. 2003, \aap, 410, 527

\bibitem[{{Bensby} {et~al.}(2014){Bensby}, {Feltzing}, \&
  {Oey}}]{2014A&A...562A..71B}
{Bensby}, T., {Feltzing}, S., \& {Oey}, M.~S. 2014, \aap, 562, A71

\bibitem[{{Bertelli} {et~al.}(2008){Bertelli}, {Girardi}, {Marigo}, \&
  {Nasi}}]{2008A&A...484..815B}
{Bertelli}, G., {Girardi}, L., {Marigo}, P., \& {Nasi}, E. 2008, \aap, 484, 815

\bibitem[{{Blackwell} \& {Lynas-Gray}(1998)}]{1998A&AS..129..505B}
{Blackwell}, D.~E. \& {Lynas-Gray}, A.~E. 1998, \aaps, 129, 505

\bibitem[{{Bobylev} {et~al.}(2011){Bobylev}, {Bajkova}, {Myll{\"a}ri}, \&
  {Valtonen}}]{2011AstL...37..550B}
{Bobylev}, V.~V., {Bajkova}, A.~T., {Myll{\"a}ri}, A., \& {Valtonen}, M. 2011,
  Astronomy Letters, 37, 550

\bibitem[{{Brown} {et~al.}(2010){Brown}, {Portegies Zwart}, \&
  {Bean}}]{2010MNRAS.407..458B}
{Brown}, A.~G.~A., {Portegies Zwart}, S.~F., \& {Bean}, J. 2010, MNRAS, 407,
  458

\bibitem[{{Bruntt} {et~al.}(2010){Bruntt}, {Bedding}, {Quirion}, {Lo Curto},
  {Carrier}, {Smalley}, {Dall}, {Arentoft}, {Bazot}, \&
  {Butler}}]{2010MNRAS.405.1907B}
{Bruntt}, H., {Bedding}, T.~R., {Quirion}, P.-O., {et~al.} 2010, \mnras, 405,
  1907

\bibitem[{{Casagrande} {et~al.}(2011){Casagrande}, {Sch{\"o}nrich}, {Asplund},
  {Cassisi}, {Ram{\'{\i}}rez}, {Mel{\'e}ndez}, {Bensby}, \&
  {Feltzing}}]{2011A&A...530A.138C}
{Casagrande}, L., {Sch{\"o}nrich}, R., {Asplund}, M., {et~al.} 2011, \aap, 530,
  A138

\bibitem[{{da Silva} {et~al.}(2006){da Silva}, {Girardi}, {Pasquini},
  {Setiawan}, {von der L{\"u}he}, {de Medeiros}, {Hatzes}, {D{\"o}llinger}, \&
  {Weiss}}]{2006A&A...458..609D}
{da Silva}, L., {Girardi}, L., {Pasquini}, L., {et~al.} 2006, \aap, 458, 609

\bibitem[{{De Silva} {et~al.}(2007{\natexlab{a}}){De Silva}, {Freeman},
  {Asplund}, {Bland-Hawthorn}, {Bessell}, \& {Collet}}]{2007AJ....133.1161D}
{De Silva}, G.~M., {Freeman}, K.~C., {Asplund}, M., {et~al.}
  2007{\natexlab{a}}, \aj, 133, 1161

\bibitem[{{De Silva} {et~al.}(2009){De Silva}, {Freeman}, \&
  {Bland-Hawthorn}}]{2009PASA...26...11D}
{De Silva}, G.~M., {Freeman}, K.~C., \& {Bland-Hawthorn}, J. 2009, \pasa, 26,
  11

\bibitem[{{De Silva} {et~al.}(2007{\natexlab{b}}){De Silva}, {Freeman},
  {Bland-Hawthorn}, {Asplund}, \& {Bessell}}]{2007AJ....133..694D}
{De Silva}, G.~M., {Freeman}, K.~C., {Bland-Hawthorn}, J., {Asplund}, M., \&
  {Bessell}, M.~S. 2007{\natexlab{b}}, \aj, 133, 694

\bibitem[{{Dekker} {et~al.}(2000){Dekker}, {D'Odorico}, {Kaufer}, {Delabre}, \&
  {Kotzlowski}}]{2000SPIE.4008..534D}
{Dekker}, H., {D'Odorico}, S., {Kaufer}, A., {Delabre}, B., \& {Kotzlowski}, H.
  2000, in Society of Photo-Optical Instrumentation Engineers (SPIE) Conference
  Series, Vol. 4008, Society of Photo-Optical Instrumentation Engineers (SPIE)
  Conference Series, ed. M.~{Iye} \& A.~F. {Moorwood}, 534--545

\bibitem[{{Demarque} {et~al.}(2004){Demarque}, {Woo}, {Kim}, \&
  {Yi}}]{2004ApJS..155..667D}
{Demarque}, P., {Woo}, J.-H., {Kim}, Y.-C., \& {Yi}, S.~K. 2004, \apjs, 155,
  667

\bibitem[{{Di Folco} {et~al.}(2004){Di Folco}, {Th{\'e}venin}, {Kervella},
  {Domiciano de Souza}, {Coud{\'e} du Foresto}, {S{\'e}gransan}, \&
  {Morel}}]{2004A&A...426..601D}
{Di Folco}, E., {Th{\'e}venin}, F., {Kervella}, P., {et~al.} 2004, \aap, 426,
  601

\bibitem[{{Feltzing} {et~al.}(2001){Feltzing}, {Holmberg}, \&
  {Hurley}}]{2001A&A...377..911F}
{Feltzing}, S., {Holmberg}, J., \& {Hurley}, J.~R. 2001, \aap, 377, 911

\bibitem[{{Ferrers}(1877)}]{1877pam..14..1}
{Ferrers}, N.~M. 1877, Q.J. Pure Appl. Math., 14, 1

\bibitem[{{Flower}(1996)}]{1996ApJ...469..355F}
{Flower}, P.~J. 1996, \apj, 469, 355

\bibitem[{{Frankowski} {et~al.}(2007){Frankowski}, {Jancart}, \&
  {Jorissen}}]{2007A&A...464..377F}
{Frankowski}, A., {Jancart}, S., \& {Jorissen}, A. 2007, \aap, 464, 377

\bibitem[{{Freeman} \& {Bland-Hawthorn}(2002)}]{2002ARA&A..40..487F}
{Freeman}, K. \& {Bland-Hawthorn}, J. 2002, \araa, 40, 487

\bibitem[{{Gray} {et~al.}(2000){Gray}, {Tycner}, \&
  {Brown}}]{2000PASP..112..328G}
{Gray}, D.~F., {Tycner}, C., \& {Brown}, K. 2000, \pasp, 112, 328

\bibitem[{{Grevesse} {et~al.}(2007){Grevesse}, {Asplund}, \&
  {Sauval}}]{2007SSRv..130..105G}
{Grevesse}, N., {Asplund}, M., \& {Sauval}, A.~J. 2007, \ssr, 130, 105

\bibitem[{{Gustafsson} {et~al.}(2008){Gustafsson}, {Edvardsson}, {Eriksson},
  {J{\o}rgensen}, {Nordlund}, \& {Plez}}]{2008A&A...486..951G}
{Gustafsson}, B., {Edvardsson}, B., {Eriksson}, K., {et~al.} 2008, \aap, 486,
  951

\bibitem[{{H{\"a}nninen} \& {Flynn}(2002)}]{2002MNRAS.337..731H}
{H{\"a}nninen}, J. \& {Flynn}, C. 2002, \mnras, 337, 731

\bibitem[{{Holmberg} {et~al.}(2009){Holmberg}, {Nordstr{\"o}m}, \&
  {Andersen}}]{2009A&A...501..941H}
{Holmberg}, J., {Nordstr{\"o}m}, B., \& {Andersen}, J. 2009, \aap, 501, 941

\bibitem[{{Jofre} {et~al.}(2013{\natexlab{a}}){Jofre}, {Heiter},
  {Blanco-Cuaresma}, \& {Soubiran}}]{2013arXiv1312.2943J}
{Jofre}, P., {Heiter}, U., {Blanco-Cuaresma}, S., \& {Soubiran}, C.
  2013{\natexlab{a}}, ArXiv e-prints

\bibitem[{{Jofre} {et~al.}(2013{\natexlab{b}}){Jofre}, {Heiter}, {Soubiran},
  {Blanco-Cuaresma}, {Pancino}, {Bergemann}, {Cantat-Gaudin}, {Gonzalez
  Hernandez}, {Hill}, {Lardo}, {de Laverny}, {Lind}, {Magrini}, {Masseron},
  {Montes}, {Mucciarelli}, {Nordlander}, {Recio-Blanco}, {Sobeck}, {Sordo},
  {Sousa}, {Tabernero}, {Vallenari}, {Van Eck}, \&
  {Worley}}]{2013arXiv1309.1099J}
{Jofre}, P., {Heiter}, U., {Soubiran}, C., {et~al.} 2013{\natexlab{b}}, ArXiv
  e-prints

\bibitem[{{J{\o}rgensen} \& {Lindegren}(2005)}]{2005A&A...436..127J}
{J{\o}rgensen}, B.~R. \& {Lindegren}, L. 2005, \aap, 436, 127

\bibitem[{{Kaufer} {et~al.}(1999){Kaufer}, {Stahl}, {Tubbesing},
  {N{\o}rregaard}, {Avila}, {Francois}, {Pasquini}, \&
  {Pizzella}}]{1999Msngr..95....8K}
{Kaufer}, A., {Stahl}, O., {Tubbesing}, S., {et~al.} 1999, The Messenger, 95, 8

\bibitem[{{Kervella} {et~al.}(2004){Kervella}, {Th{\'e}venin}, {Di Folco}, \&
  {S{\'e}gransan}}]{2004A&A...426..297K}
{Kervella}, P., {Th{\'e}venin}, F., {Di Folco}, E., \& {S{\'e}gransan}, D.
  2004, \aap, 426, 297

\bibitem[{{Kroupa}(2001)}]{2001MNRAS.322..231K}
{Kroupa}, P. 2001, \mnras, 322, 231

\bibitem[{{Lada} \& {Lada}(2003)}]{2003ARA&A..41...57L}
{Lada}, C.~J. \& {Lada}, E.~A. 2003, \araa, 41, 57

\bibitem[{{Lallement} {et~al.}(2003){Lallement}, {Welsh}, {Vergely}, {Crifo},
  \& {Sfeir}}]{2003A&A...411..447L}
{Lallement}, R., {Welsh}, B.~Y., {Vergely}, J.~L., {Crifo}, F., \& {Sfeir}, D.
  2003, \aap, 411, 447

\bibitem[{{Mamajek} \& {Hillenbrand}(2008)}]{2008ApJ...687.1264M}
{Mamajek}, E.~E. \& {Hillenbrand}, L.~A. 2008, \apj, 687, 1264

\bibitem[{{Mart{\'{\i}}nez-Barbosa} {et~al.}(2014){Mart{\'{\i}}nez-Barbosa},
  {Brown}, \& {Portegies Zwart}}]{2014arXiv1410.2238M}
{Mart{\'{\i}}nez-Barbosa}, C.~A., {Brown}, A.~G.~A., \& {Portegies Zwart}, S.
  2014, ArXiv e-prints

\bibitem[{{Mayor} {et~al.}(2003){Mayor}, {Pepe}, {Queloz}, {Bouchy},
  {Rupprecht}, {Lo Curto}, {Avila}, {Benz}, {Bertaux}, {Bonfils}, {Dall},
  {Dekker}, {Delabre}, {Eckert}, {Fleury}, {Gilliotte}, {Gojak}, {Guzman},
  {Kohler}, {Lizon}, {Longinotti}, {Lovis}, {Megevand}, {Pasquini}, {Reyes},
  {Sivan}, {Sosnowska}, {Soto}, {Udry}, {van Kesteren}, {Weber}, \&
  {Weilenmann}}]{2003Msngr.114...20M}
{Mayor}, M., {Pepe}, F., {Queloz}, D., {et~al.} 2003, The Messenger, 114, 20

\bibitem[{{Mishurov} \& {Acharova}(2011)}]{2011MNRAS.412.1771M}
{Mishurov}, Y.~N. \& {Acharova}, I.~A. 2011, \mnras, 412, 1771

\bibitem[{{Mitschang} {et~al.}(2013){Mitschang}, {De Silva}, {Sharma}, \&
  {Zucker}}]{2013MNRAS.428.2321M}
{Mitschang}, A.~W., {De Silva}, G., {Sharma}, S., \& {Zucker}, D.~B. 2013,
  \mnras, 428, 2321

\bibitem[{{Mitschang} {et~al.}(2014){Mitschang}, {De Silva}, {Zucker},
  {Anguiano}, {Bensby}, \& {Feltzing}}]{2014MNRAS.438.2753M}
{Mitschang}, A.~W., {De Silva}, G., {Zucker}, D.~B., {et~al.} 2014, \mnras,
  438, 2753

\bibitem[{{Nieva} \& {Przybilla}(2012)}]{2012A&A...539A.143N}
{Nieva}, M.-F. \& {Przybilla}, N. 2012, \aap, 539, A143

\bibitem[{{Nissen} {et~al.}(2014){Nissen}, {Chen}, {Carigi}, {Schuster}, \&
  {Zhao}}]{2014A&A...568A..25N}
{Nissen}, P.~E., {Chen}, Y.~Q., {Carigi}, L., {Schuster}, W.~J., \& {Zhao}, G.
  2014, \aap, 568, A25

\bibitem[{{Nordstr{\"o}m} {et~al.}(2004){Nordstr{\"o}m}, {Mayor}, {Andersen},
  {Holmberg}, {Pont}, {J{\o}rgensen}, {Olsen}, {Udry}, \&
  {Mowlavi}}]{2004A&A...418..989N}
{Nordstr{\"o}m}, B., {Mayor}, M., {Andersen}, J., {et~al.} 2004, \aap, 418, 989

\bibitem[{{North} {et~al.}(2007){North}, {Davis}, {Bedding}, {Ireland},
  {Jacob}, {O'Byrne}, {Owens}, {Robertson}, {Tango}, \&
  {Tuthill}}]{2007MNRAS.380L..80N}
{North}, J.~R., {Davis}, J., {Bedding}, T.~R., {et~al.} 2007, \mnras, 380, L80

\bibitem[{{Olsen}(1983)}]{1983A&AS...54...55O}
{Olsen}, E.~H. 1983, \aaps, 54, 55

\bibitem[{{Olsen}(1984)}]{1984A&AS...57..443O}
{Olsen}, E.~H. 1984, \aaps, 57, 443

\bibitem[{{Olsen}(1994)}]{1994A&AS..106..257O}
{Olsen}, E.~H. 1994, \aaps, 106, 257

\bibitem[{{Pancino} {et~al.}(2010){Pancino}, {Carrera}, {Rossetti}, \&
  {Gallart}}]{2010A&A...511A..56P}
{Pancino}, E., {Carrera}, R., {Rossetti}, E., \& {Gallart}, C. 2010, \aap, 511,
  A56

\bibitem[{{Pavlenko} {et~al.}(2012){Pavlenko}, {Jenkins}, {Jones}, {Ivanyuk},
  \& {Pinfield}}]{2012MNRAS.422..542P}
{Pavlenko}, Y.~V., {Jenkins}, J.~S., {Jones}, H.~R.~A., {Ivanyuk}, O., \&
  {Pinfield}, D.~J. 2012, \mnras, 422, 542

\bibitem[{{Pelupessy} {et~al.}(2012){Pelupessy}, {J{\"a}nes}, \& {Portegies
  Zwart}}]{2012NewA...17..711P}
{Pelupessy}, F.~I., {J{\"a}nes}, J., \& {Portegies Zwart}, S. 2012, \na, 17,
  711

\bibitem[{{Perryman} {et~al.}(1998){Perryman}, {Brown}, {Lebreton}, {Gomez},
  {Turon}, {Cayrel de Strobel}, {Mermilliod}, {Robichon}, {Kovalevsky}, \&
  {Crifo}}]{1998A&A...331...81P}
{Perryman}, M.~A.~C., {Brown}, A.~G.~A., {Lebreton}, Y., {et~al.} 1998, \aap,
  331, 81

\bibitem[{{Pichardo} {et~al.}(2004){Pichardo}, {Martos}, \&
  {Moreno}}]{2004ApJ...609..144P}
{Pichardo}, B., {Martos}, M., \& {Moreno}, E. 2004, \apj, 609, 144

\bibitem[{{Pichardo} {et~al.}(2012){Pichardo}, {Moreno}, {Allen}, {Bedin},
  {Bellini}, \& {Pasquini}}]{2012AJ....143...73P}
{Pichardo}, B., {Moreno}, E., {Allen}, C., {et~al.} 2012, \aj, 143, 73

\bibitem[{{Plummer}(1911)}]{1911MNRAS..71..460P}
{Plummer}, H.~C. 1911, \mnras, 71, 460

\bibitem[{{Portegies Zwart} {et~al.}(2013){Portegies Zwart}, {McMillan}, {van
  Elteren}, {Pelupessy}, \& {de Vries}}]{2013CoPhC.183..456P}
{Portegies Zwart}, S., {McMillan}, S.~L.~W., {van Elteren}, E., {Pelupessy},
  I., \& {de Vries}, N. 2013, Computer Physics Communications, 183, 456

\bibitem[{{Portegies Zwart}(2009)}]{2009ApJ...696L..13P}
{Portegies Zwart}, S.~F. 2009, ApJ, 696, L13

\bibitem[{{Portegies Zwart} \& {Verbunt}(1996)}]{1996A&A...309..179P}
{Portegies Zwart}, S.~F. \& {Verbunt}, F. 1996, \aap, 309, 179

\bibitem[{{Radick} {et~al.}(1987){Radick}, {Thompson}, {Lockwood}, {Duncan}, \&
  {Baggett}}]{1987ApJ...321..459R}
{Radick}, R.~R., {Thompson}, D.~T., {Lockwood}, G.~W., {Duncan}, D.~K., \&
  {Baggett}, W.~E. 1987, \apj, 321, 459

\bibitem[{{Ram{\'{\i}}rez} {et~al.}(2014){Ram{\'{\i}}rez}, {Bajkova},
  {Bobylev}, {Roederer}, {Lambert}, {Endl}, {Cochran}, {MacQueen}, \&
  {Wittenmyer}}]{2014ApJ...787..154R}
{Ram{\'{\i}}rez}, I., {Bajkova}, A.~T., {Bobylev}, V.~V., {et~al.} 2014, \apj,
  787, 154

\bibitem[{{Reiners} \& {Schmitt}(2003)}]{2003A&A...398..647R}
{Reiners}, A. \& {Schmitt}, J.~H.~M.~M. 2003, \aap, 398, 647

\bibitem[{{Romero-G{\'o}mez} {et~al.}(2011){Romero-G{\'o}mez}, {Athanassoula},
  {Antoja}, \& {Figueras}}]{2011MNRAS.418.1176R}
{Romero-G{\'o}mez}, M., {Athanassoula}, E., {Antoja}, T., \& {Figueras}, F.
  2011, \mnras, 418, 1176

\bibitem[{{Saar} \& {Osten}(1997)}]{1997MNRAS.284..803S}
{Saar}, S.~H. \& {Osten}, R.~A. 1997, \mnras, 284, 803

\bibitem[{{Santos} {et~al.}(2004){Santos}, {Israelian}, \&
  {Mayor}}]{2004A&A...415.1153S}
{Santos}, N.~C., {Israelian}, G., \& {Mayor}, M. 2004, \aap, 415, 1153

\bibitem[{{Sellwood}(2014)}]{2014RvMP...86....1S}
{Sellwood}, J.~A. 2014, Reviews of Modern Physics, 86, 1

\bibitem[{{Sellwood} \& {Binney}(2002)}]{2002MNRAS.336..785S}
{Sellwood}, J.~A. \& {Binney}, J.~J. 2002, \mnras, 336, 785

\bibitem[{{Soderblom}(2010)}]{2010ARA&A..48..581S}
{Soderblom}, D.~R. 2010, \araa, 48, 581

\bibitem[{{Takeda} {et~al.}(2007){Takeda}, {Ford}, {Sills}, {Rasio}, {Fischer},
  \& {Valenti}}]{2007ApJS..168..297T}
{Takeda}, G., {Ford}, E.~B., {Sills}, A., {et~al.} 2007, \apjs, 168, 297

\bibitem[{{Toonen} {et~al.}(2012){Toonen}, {Nelemans}, \& {Portegies
  Zwart}}]{2012A&A...546A..70T}
{Toonen}, S., {Nelemans}, G., \& {Portegies Zwart}, S. 2012, \aap, 546, A70

\bibitem[{{Torres}(2010)}]{2010AJ....140.1158T}
{Torres}, G. 2010, \aj, 140, 1158

\bibitem[{{Valenti} \& {Fischer}(2005)}]{2005ApJS..159..141V}
{Valenti}, J.~A. \& {Fischer}, D.~A. 2005, \apjs, 159, 141

\bibitem[{{Valenti} \& {Piskunov}(1996)}]{1996A&AS..118..595V}
{Valenti}, J.~A. \& {Piskunov}, N. 1996, \aaps, 118, 595

\bibitem[{{van Leeuwen}(2007)}]{2007A&A...474..653V}
{van Leeuwen}, F. 2007, \aap, 474, 653

\bibitem[{{Volz} {et~al.}(1996){Volz}, {Majerus}, {Liebel}, {Schmitt}, \&
  {Schmoranzer}}]{1996PhRvL..76.2862V}
{Volz}, U., {Majerus}, M., {Liebel}, H., {Schmitt}, A., \& {Schmoranzer}, H.
  1996, Physical Review Letters, 76, 2862

\bibitem[{{Wielen} {et~al.}(1996){Wielen}, {Fuchs}, \&
  {Dettbarn}}]{1996A&A...314..438W}
{Wielen}, R., {Fuchs}, B., \& {Dettbarn}, C. 1996, \aap, 314, 438

\bibitem[{{Yi} {et~al.}(2003){Yi}, {Kim}, \& {Demarque}}]{2003ApJS..144..259Y}
{Yi}, S.~K., {Kim}, Y.-C., \& {Demarque}, P. 2003, \apjs, 144, 259

\end{thebibliography}

\end{document}